\documentclass[iop,apj,twocolappendix]{emulateapj}
\usepackage{apjfonts}
 \usepackage{color}

\newcommand{\beq}{\begin{equation}}
\newcommand{\eeq}{\end{equation}}
\newcommand{\beqn}{\begin{eqnarray}}
\newcommand{\eeqn}{\end{eqnarray}}
\newcommand{\pd}{\partial}

\newcommand{\md}{{\rm mid}}
\newcommand{\ideal}{{\rm ideal}}
\newcommand{\res}{{\rm res}}
\newcommand{\rhodvtwo}{\bracket{\rho}\bracket{\delta v^2}}
\newcommand{\rhodvztwo}{\bracket{\rho}\bracket{\delta v_z^2}}
\newcommand{\alphatot}{\alpha}
\newcommand{\alphacore}{\alpha_{\rm core}}
\newcommand{\alphaatm}{\alpha_{\rm atm}}
\newcommand{\ovl}[1]{ {\overline{#1}} }

\newcommand{\bracket}[1]{\langle #1 \rangle}

\newcommand{\eqref}[1]{(\ref{#1})}

\newcommand{\dfrac}[2]{ {\displaystyle\frac{#1}{#2}} }
\newcommand{\pfrac}[2]{ \biggl(\dfrac{#1}{#2}\biggr) }

\renewcommand{\leq}{\leqslant}
\renewcommand{\geq}{\geqslant}

\shorttitle{MODELING MRI TURBULENCE IN PROTOPLANETARY DISKS WITH DEAD ZONES}
\shortauthors{OKUZUMI \& HIROSE}
\slugcomment{ApJ Accepted}

\begin{document}
\title{Modeling Magnetorotational Turbulence in Protoplanetary Disks with Dead Zones}
\author{Satoshi Okuzumi}
\affil{Department of Physics, Nagoya University, Nagoya, Aichi 464-8602, Japan; okuzumi@nagoya-u.jp}
\and
\author{Shigenobu Hirose}
\affil{Institute for Research on Earth Evolution, JAMSTEC, Yokohama, Kanagawa 236-0001, Japan}

\begin{abstract}
Turbulence driven by magnetorotational instability (MRI) crucially affects the evolution of solid bodies in protoplanetary disks. On the other hand, small dust particles stabilize MRI by capturing ionized gas particles needed for the coupling of the gas and magnetic fields. To provide an empirical basis for modeling the coevolution of dust and MRI, we perform three-dimensional, ohmic-resistive MHD simulations of a vertically stratified shearing box with an MRI-inactive ``dead zone'' of various sizes and with a net vertical magnetic flux of various strengths. We find that the vertical structure of turbulence is well characterized by the vertical magnetic flux and three critical heights derived from the linear analysis of MRI in a stratified disk. In particular, the turbulent structure depends on the resistivity profile only through the critical heights and is insensitive to the details of the resistivity profile. We discover scaling relations between the amplitudes of various turbulent quantities (velocity dispersion, density fluctuation, vertical diffusion coefficient, and outflow mass flux) and vertically integrated accretion stresses. We also obtain empirical formulae for the integrated accretion stresses as a function of the vertical magnetic flux and the critical heights. These empirical relations allow to predict the vertical turbulent structure of a protoplanetary disk for a given strength of the magnetic flux and a given resistivity profile. 
 \end{abstract}
\maketitle

\section{Introduction}
Planets are believed to form in protoplanetary gas disks. 
The standard scenario for planet formation consists of the following steps. 
Initially, submicron-sized dust grains grow into kilometer-sized planetesimals 
by collisional sticking and/or gravitational instability \citep{S69,GW73,WC93}.
Planetesimals undergo further growth toward Moon-sized protoplanets 
through mutual collision assisted by gravitational interaction \citep{WS89}.
Accretion of the disk gas onto protoplanets leads to the formation of gas giants
\citep{M80,P+96}, 
while terrestrial planets form through the giant impacts of protoplanets 
after the gas disk disperses by viscous accretion onto the central star and other effects \citep{CW98}.

Turbulence in protoplanetary disks plays a decisive role on planet formation as well as on disk dispersal.
The impact of turbulence is particularly strong on the formation of 
planetesimals since the frictional coupling of gas and dust particles governs the process.
Classically, planetesimal formation has been attributed to 
the collapse of a dust sedimentary layer by self-gravity \citep{S69,GW73} 
and/or the collisional growth of dust grains \citep{WC93}.
The presence of strong turbulence is preferable for dust growth when
the dust particles is so small that Coulomb repulsion is effective \citep{O09,OTTS11b}.
However, strong turbulence acts against the growth of macroscopic dust aggregates 
since it makes their collision disruptive \citep{W84,J+08}. 
Furthermore, turbulence causes the diffusion of a dust sedimentary layer,
making planetesimal formation via gravitational instability difficult as well \citep{W84}.
Turbulence is also known to concentrate dust particles of particular sizes, 
but its relevance to planetesimal formation via gravitational instability 
is still under debate \citep{C+01,CHS08,CHB10,P+11}.
More recently, it has been suggested that two-fluid instability of gas and dust 
can produce dust clumps with density high enough for gravitational collapse, 
but successful dust coagulation to macroscopic sizes seems to be still required 
for this mechanism to become viable \citep{YG05,JY07,J+07,BS10}. 
Besides, turbulence also affects planetesimal growth as turbulent density fluctuations
gravitationally interact with planetesimals and can raise their random velocities 
above the escape velocity \citep{IGM08,NG10}.
The fluctuating gravitational field can even cause 
random orbital migration of protoplanets \citep{LSA04,NP04}.
Thus, to understand the growth of solid bodies in various stages,
it is essential to know the strength and spatial distribution of disk turbulence.

Interestingly, the evolution of solid bodies is not only affected but also affects disk turbulence.
The most viable mechanism for generating disk turbulence 
is the magnetorotational instability (MRI; \citealt*{BH91}).
This instability has its origin in the interaction between the gas disk and magnetic fields,
and therefore requires a sufficiently high ionization degree to operate. 
Importantly, whether the MRI operates or not in each location of the disk 
is strongly dependent on the amount of small dust grains 
because they efficiently capture ionized gas particles 
and thus reduce the ionization degree \citep{SMUN00,IN06a,O09}.
This implies that dust and MRI-driven turbulence affect each other and thus 
evolve simultaneously. 

The purpose of this study is to present an empirical basis 
for studying the coevolution of solid particles and MRI-driven turbulence.
It is computationally intensive 
to simulate the evolution of dust and MRI-driven turbulence simultaneously,
since the evolutionary timescale of solid bodies is generally much longer than 
the dynamical timescale of the turbulence.
For example, turbulent eddies grow and decay on a timescale of one orbital period \citep[e.g.,][]{FP06}, 
while dust particles grow to macroscopic sizes and settle to the midplane  
spending 100--1000 orbital periods \citep[e.g.,][]{NNH81,DD05,BDH08}. 
However, this also means that MRI-driven turbulence can be regarded
as quasi-steady in each evolutionary stage of dust evolution.
Motivated by this fact, we restrict ourselves to time-independent ohmic resistivity,
but instead focus on how the quasi-steady structure of 
turbulence  depends on the vertical profile of the resistivity. 

To characterize the vertical structure of MRI-driven turbulence, we perform
a number of three-dimensional MHD simulations of local stratified disks 
including resistivity and nonzero net vertical magnetic flux.
Inclusion of a nonzero net vertical flux is important 
as it determines the saturation level of turbulence 
\citep[][see also our Section~\ref{sec:4}]{HGB95,SITS04,SI09}.
Similar simulations have been done in a number of previous studies 
\citep[e.g.,][]{MS00,SI09,OM09,SMI10,TCS10,GNT11,SHB11,HT11}.
One important difference between our study and previous ones is that 
we focus on general dependence of the saturated turbulent state on the model parameters
such as the resistivity and net magnetic vertical flux.

Our modeling of MRI-driven turbulence follows two steps.
In the first step, we seek scaling laws giving the relations among turbulent quantities.
We express the relations as a function of the vertically integrated accretion stress,
which is the quantity that determines the rate at which turbulent energy is extracted 
from the differential rotation \citep{BP99}.
As we will see, excellent scaling relations are obtained if we divide the integrated stress 
into two components that characterize the contributions from different regions in the stratified disk  
(which we will call the ``{\it disk core}'' and ``{\it atmosphere}'') 
In the second step, we find out empirical formulae that predict the vertically integrated stresses
as a function of the resistivity profile and vertical magnetic flux.  

The plan of this paper is as follows. 
In Section \ref{sec:2}, we describe the method and setup used in our MHD simulations.
In Section \ref{sec:3}, we introduce ``critical heights'' derived from  
the linear analysis of MRI in stratified disks.
As we will see later, these critical heights are useful to characterize the turbulent 
structure  observed in our simulations.
We present our simulation results in Section \ref{sec:4}, 
and obtain scaling relations and predictor functions 
for the quasi-steady state of turbulence in Section \ref{sec:5}.
In Section \ref{sec:6}, we simulate dust settling in a dead zone 
to model the diffusion coefficient for small particles as a function of height.
Effects of numerical resolutions on our simulation results are discussed in Section \ref{sec:7}.
Our findings are summarized in Section \ref{sec:8}.   

\section{Simulation setup and Method }\label{sec:2}
In this section, we describe the setup and method adopted 
in our stratified resistive MHD simulations.

\subsection{Setup }\label{sec:2.1}
Our MHD simulations adopt the shearing box approximation \citep{HGB95}.
We consider a small patch of disk centered on the midplane of 
an arbitrary distance from the central star,
and model it as a stratified shearing box corotating 
with the angular speed $\Omega$ at the domain center.
We use the Cartesian coordinate system $(x, y, z)$, 
where $x$, $y$, and $z$ stand for the radial, azimuthal, and vertical 
distance from the domain center.
In addition, we assume that the gas is isothermal throughout the box; 
thus, the sound velocity $c_s$ of the gas is constant in both time and space.

\subsubsection{Initial Conditions}\label{sec:2.1.1}
For the initial condition, we assume that the gas disk is initially
in hydrostatic equilibrium and is threaded by uniform vertical magnetic field $B_{z0}$.
The assumption of the hydrostatic equilibrium 
leas to the initial gas density profile
\beq
\rho = \rho_0 \exp\left(-\frac{z^2}{2h^2}\right),
\label{eq:rho}
\eeq
where $\rho_0$ is the initial gas density at the midplane and 
\beq
h = \frac{c_s}{\Omega}
\eeq
is the pressure scale height.\footnote{\label{foot}Note that 
the ``gas scale height'' is often defined as $H = \sqrt{2}h = \sqrt{2}c_s/\Omega$
in the literature on stratified MHD simulations.
} 
The ratio between the initial midplane gas pressure $P_0 = \rho_0 c_s^2$ 
and the initial magnetic pressure $B_{z0}^2/8\pi$
defines the initial plasma beta
\beq
\beta_{z0} \equiv \frac{8\pi \rho_0 c_s^2}{B_{z0}^2}.
\eeq
In this paper, the strength of the initial magnetic flux will be referred by $\beta_{z0}^{-1}$
rather than $B_{z0}$.

\subsubsection{Resistivity Profile}\label{sec:2.1.2}
The main purpose of this study is to see how the turbulence depends on 
the vertical profile of the ohmic resistivity.
We adopt a simple analytic resistivity profile based on the following consideration.
For fixed temperature, the resistivity is inversely proportional 
to the ionization degree \citep{BB94}.
In protoplanetary disks, the ionization degree at each location is determined by 
the balance of ionization (by, e.g., cosmic rays and X-rays) and recombination
(in the gas phase and on grain surfaces).
Detailed structure of the resistivity profile depends on what processes dominate the
ionization and recombination.
However, a general tendency is that the ionization degree 
decreases toward the midplane of the disk, 
because the ionization rate is lower as the column depth is greater
and because the recombination rate is higher as the gas density is higher 
\citep[see, e.g.,][]{SMUN00}.
Based on this fact, we give the resistivity profile $\eta(z)$ 
such that $\eta$ increases as $z$ decreases.
To be more specific, we adopt the following resistivity profile
\beq
\eta = \eta_\md \exp\left(-\frac{z^2}{2h_\eta^2}\right),
\label{eq:eta}
\eeq
where $\eta_\md$ is the resistivity at the midplane and $h_\eta$ 
is the scale height of $\eta$.

\begin{deluxetable}{lllcccccc}
\tablecaption{Model Parameters and Initial Critical Heights}
\tablecolumns{8}
\tablewidth{0pt}
\tablehead{
\colhead{Model} & \colhead{$\dfrac{\eta_\md}{c_sh}$} &  
\colhead{$\dfrac{h^2}{h_\eta^2}$} &
\colhead{$\beta_{z0}$}  & \colhead{$\log(\Lambda_0)$} &
 \colhead{$\dfrac{h_{\ideal,0}}{h}$} & 
 \colhead{$\dfrac{h_{\Lambda,0}}{h}$}  &
  \colhead{$\dfrac{h_{\res,0}}{h}$}
}
\startdata
Ideal & $0$ & \nodata & $3 \times 10^5$ & $\infty$ & $4.1$ & $0.0$ & $0.0$ \\ 
X0 & $0.02$ & \phn$0.3$ & $3 \times 10^5$ & $-3.5$ & $4.1$ & $3.5$ & $2.8$ \\ 
X1 & $0.02$ & \phn$1.0$ & $3 \times 10^5$ & $-3.5$ & $4.1$ & $2.8$ & $2.2$ \\ 
X2 & $0.02$ & \phn$4.0$ & $3 \times 10^5$ & $-3.5$ & $4.1$ & $1.8$ & $1.3$ \\ 
X3 & $0.02$ & $20.0$ & $3 \times 10^5$ & $-3.5$ & $4.1$ & $0.9$ & $0.6$ \\ 
Y1 & $0.0004$ & \phn$0.0$ & $3 \times 10^5$ & $-1.8$ & $4.1$ & $2.9$ & $0.0$ \\ 
Y2 & $0.0004$ & \phn$1.5$ & $3 \times 10^5$ & $-1.8$ & $4.1$ & $1.8$ & $0.0$ \\ 
Y3 & $0.0004$ & \phn$9.5$ & $3 \times 10^5$ & $-1.8$ & $4.1$ & $0.9$ & $0.0$ \\ 
Y4 & $0.0004$ & $40.0$ & $3 \times 10^5$ & $-1.8$ & $4.1$ & $0.4$ & $0.0$ \\ 
W1 & $1.0$ & \phn$2.0$ & $3 \times 10^5$ & $-5.2$ & $4.1$ & $2.8$ & $2.4$ \\ 
W2 & $1.0$ & \phn$6.4$ & $3 \times 10^5$ & $-5.2$ & $4.1$ & $1.8$ & $1.5$ \\ 
W3 & $1.0$ & $30.0$ & $3 \times 10^5$ & $-5.2$ & $4.1$ & $0.9$ & $0.7$ \\ 
X1a & $0.02$ & \phn$1.0$ & $1 \times 10^7$ & $-5.0$ & $4.8$ & $3.4$ & $2.7$ \\ 
X1b & $0.02$ & \phn$1.0$ & $3 \times 10^6$ & $-4.5$ & $4.6$ & $3.2$ & $2.5$ \\ 
X1c & $0.02$ & \phn$1.0$ & $1 \times 10^5$ & $-3.0$ & $3.8$ & $2.6$ & $1.9$ \\ 
X1d & $0.02$ & \phn$1.0$ & $3 \times 10^4$ & $-2.5$ & $3.4$ & $2.4$ & $1.4$ \\ 
FS03L & $0.01$ & \nodata & $3 \times 10^5$ & $-3.2$ & $4.1$ & $1.3$ & $0.6$ 
\enddata
\label{tab:param}
\end{deluxetable}

Equation~\eqref{eq:eta} satisfies the important property 
of realistic resistivity profiles mentioned above.
Furthermore, as shown in Appendix, Equation~\eqref{eq:eta} exactly reproduces 
the vertical resistivity profile of a disk in some limited cases.
However, it will be useful to examine 
possible influences of limiting $\eta$ to Equation~\eqref{eq:eta}.
In order to do that, we also consider a resistive profile used in \citet{FS03},
\beq
\eta_{\rm FS03} = \eta_0 \exp\left( -\frac{z^2}{4h^2}\right) 
\exp\left(\frac{\Sigma_0}{\Sigma_{\rm CR}}\frac{1}{2\sqrt{2\pi}}
\int_{|z|/h}^\infty e^{-z'^2/2} dz'\right),
\label{eq:eta_FS03}
\eeq
which is characterized by two parameters $\eta_0$ 
and $\Sigma_0/\Sigma_{\rm CR}$ (see Equation~(10) of \citealt*{FS03}).
Physically, Equation~\eqref{eq:eta_FS03} corresponds to the resistivity profile 
when the ionization degree is determined by 
the balance between cosmic-ray ionization and gas-phase recombination (see also Appendix).

We construct 17 simulation models using Equations~\eqref{eq:eta} and \eqref{eq:eta_FS03}.
16 models are constructed from Equation~\eqref{eq:eta} 
with various sets of the parameters ($\eta_\md$, $h_\eta$, $\beta_{z0}$).
Table~\ref{tab:param} lists the parameters adopted in each run.
The model ``Ideal'' assumes zero resistivity throughout the simulation box.
Models X0--X3 are defined by the same value of $\eta_\md$ but different values of $h_\eta$.
The difference between X, Y, and W models is in the value of $\eta_\md$.
The initial plasma beta is taken to be $3\times 10^5$ in all models except X1a--X1d.
In addition, we construct one model from Equation~\eqref{eq:eta_FS03}
with $\eta_{\rm mid} = \eta_0\exp(\Sigma_0/4\Sigma_{\rm CR}) = 0.01$ 
and $\Sigma_0/\Sigma_{\rm CR} = 34.7$.
This set of parameters corresponds to the ``larger dead zone'' model of \citet{FS03} 
defined by ${\rm Re}_{M,\md} = 100$ and ${\rm Re}_{M,z=2\sqrt{2}h} = 5.6\times 10^6$,
where ${\rm Re}_{M} \equiv c_sh/\eta$ is the magnetic Reynolds number
with the typical length and velocity  set to be $h$ and $c_s$, respectively.
We refer to the model with these parameters as model FS03L.
The initial plasma beta in the FS03L model is taken to be $3\times 10^5$. 
Note that our FS03L run is not exactly the same as the larger dead zone run of \citet{FS03}
since they assumed zero net vertical magnetic flux.

\subsection{Method}\label{sec:2.2}
We solve the equations of the isothermal resistive MHD using the ZEUS code \citep{SN92}.
The domain is a box of size $2\sqrt{2}h\times 8\sqrt{2}h \times 10\sqrt{2}h$
along the radial, azimuthal, and vertical directions, divided into $40 \times 80 \times 200$ grid cells.
For all runs, the radial and azimuthal boundary conditions are
taken to be shearing-periodic and periodic, respectively.
Therefore, the vertical magnetic flux is a conserved quantity of our simulations.
The vertical boundary condition for all runs except X1d is the standard ZEUS 
outflow condition, where the fields on the 
boundaries are computed from the extrapolated electromotive forces;
only for run X1d, we assume for numerical stability that the magnetic fields are vertical 
on the top and bottom boundaries as done by \citet{FKK10}.
We have checked that the results hardly depend on the choice 
of the two types of boundary conditions.
Note that the radial and azimuthal components of the mean magnetic fields 
are not conserved due to the outflow boundary condition.
We also added a small artificial resistivity near the boundaries for numerical stability \citep{HKB09}. 
A density floor of $10^{-5}\rho_0$ is
applied to prevent high Alfv\'{e}n speeds from halting the calculations.

\section{Characteristic Wavelengths and Critical Heights}\label{sec:3}
As we will see in the following sections, it is useful to analyze the simulation results
using the knowledge obtained from the linear analysis of MRI. 
In this subsection, we introduce several quantities that characterize 
the linear evolution of MRI.

\subsection{Characteristic Wavelengths of MRI}\label{sec:3.1}
According to the local linear analysis of MRI including ohmic resistivity \citep{SM99}, 
the wavelength of the most unstable MRI mode can be approximately expressed as
\beq
\lambda_{\rm local} \approx \max\{\lambda_\ideal,\lambda_\res\},
\label{eq:lambda_local}
\eeq
where 
\beq
\lambda_\ideal \equiv 2\pi \frac{v_{Az}}{\Omega}
\label{eq:lambda_ideal}
\eeq
and 
\beq
\lambda_\res \equiv 2\pi \frac{\eta}{v_{Az}}
\label{eq:lambda_res}
\eeq
are the characteristic wavelengths of MRI modes in the ideal and resistive MHD limits, respectively,
and $v_{Az} = B_z/\sqrt{4\pi \rho}$ is the vertical component of the Alfv\'{e}n velocity.
Equation~\eqref{eq:lambda_local} can be written as  
$\lambda_{\rm local} \approx \lambda_\ideal \max\{1,\Lambda^{-1}\}$, where
$\Lambda$ is the Elsasser number defined by
\beq
\Lambda \equiv \frac{v_{Az}^2}{\eta \Omega}. 
\label{eq:Lambda}
\eeq
The Elsasser number determines the growth rate of the MRI.
If $\Lambda > 1$, ohmic diffusion does not affect the most unstable mode 
lying at $\lambda = \lambda_\ideal$, and the local instability occurs rapidly 
at a wavelength $\lambda_{\rm local} \approx \lambda_\ideal$ and at a rate $\approx \Omega$.
If $\Lambda < 1$, ohmic diffusion stabilizes the most unstable mode, 
and the local instability occurs at a longer wavelength 
$\lambda_{\rm local} \approx \lambda_\res = \Lambda^{-1}\lambda_\ideal$
and at a slower rate $\approx \Lambda^{-1}\Omega$. 
We will refer to the former case as the ``ideal MRI,'' and to the latter case
as the ``resistive MRI.''

\begin{figure}
\epsscale{1}
\plotone{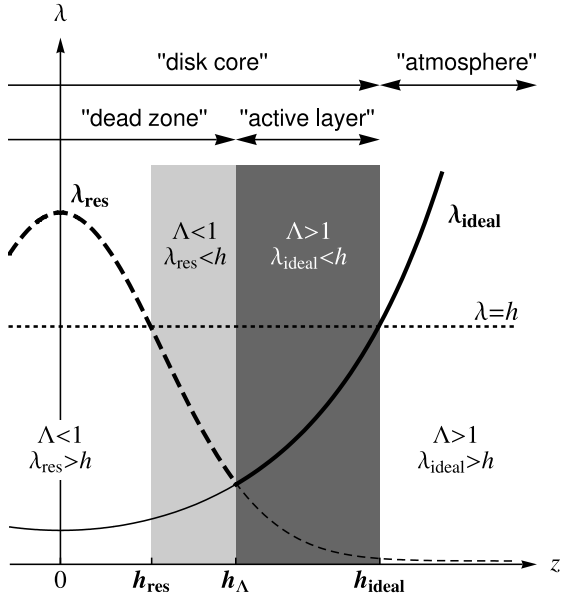}
\caption{Schematic illustration showing the vertical structure of
a stratified protoplanetary disk with vertical magnetic fields.
The horizontal axis shows the distance $z$ from the midplane, 
while the vertical axis shows the characteristic wavelengths 
$\lambda_\ideal$ (solid curve) and $\lambda_\res$ (dashed curve) of MRI at each $z$
as well as the gas scale height $h$ (dotted line).
The set of three ratios $\lambda_\ideal/h$, $\lambda_\res/h$, 
and $\lambda_\ideal/\lambda_\res \equiv \Lambda$ defines four layers.
At $|z|> h_\ideal$ ($\Lambda >1$ and $\lambda_\ideal>h$), MRI is stabilized due to 
the weak gas pressure compared to the magnetic tension.
At $h_\Lambda < |z| < h_\ideal$ ($\Lambda >1$ and $\lambda_\ideal<h$; dark gray region),
MRI operates without affected by ohmic dissipation.
At $h_\res < |z| < h_\Lambda$ ($\Lambda <1$ and $\lambda_\res<h$; light gray region),
ohmic dissipation is effective and MRI operates only weakly.
At $|z| < h_\res$ ($\Lambda <1$ and $\lambda_\res>h$),
ohmic dissipation perfectly stabilizes MRI. 
We refer to the regions $|z|<h_\ideal$ and $|z|>h_\ideal$ 
as the ``disk core'' and ``atmosphere,'' respectively.
}
\label{fig:vert}
\end{figure}
Figure~\ref{fig:vert} schematically illustrates 
how $\lambda_\ideal$ and $\lambda_\res$ vary with height $|z|$. 
In general, $\lambda_\ideal$ grows toward higher $|z|$ 
because the Alfv\'{e}n speed $v_{Az}$ increases as the density decreases. 
By contrast, $\lambda_\res$ grows toward lower $|z|$ because 
$\lambda_\res$ is inversely proportional to $v_{Az}$ and 
because $\eta$ increases with decreasing $|z|$ 
(see the discussion in Section~\ref{sec:2.1.2}).  

The global instability of a stratified disk can be described 
in terms of the local analysis.
As shown by \citet{SM99}, the gas motion at height $z$ is unstable
if the local unstable wavelength $\lambda_{\rm local}$ 
is shorter than the scale height of the disk, i.e., 
\beq
\max\{\lambda_\ideal,\lambda_\res\} \la h.
\label{eq:global}
\eeq

\subsection{Critical Heights}\label{sec:3.2}
With the global instability criterion (Equation~\eqref{eq:global}) 
together with the vertical dependence of $\lambda_\ideal$ and $\lambda_\res$,
we can define three different critical heights for a stratified disk.

\begin{enumerate}
\item 
The first one is $h_\ideal$ defined by 
\beq
\lambda_\ideal({z = h_\ideal}) = h,
\label{eq:h_ideal}
\eeq
or equivalently, $\beta_{z}(z = h_\ideal) = 8\pi^2$, 
where $\beta_z(z) = 8\pi\rho(z) c_s^2/B_z^2(z)$.
At $|z| \ga h_\ideal$ ($\beta_{z}\la 8\pi^2 $), 
MRI does not operate because the wavelengths of the unstable modes
exceed the disk thickness $\sim h$ \citep{SM99}.
We refer to the region $|z| \geq h_\ideal$ as the ``atmosphere'' 
and to the region $|z| \leq h_\ideal$ as the ``disk core.''
\item
The second one is $h_\Lambda$ defined by
\beq
\Lambda (z={h_\Lambda}) = 1,
\label{eq:h_Lambda}
\eeq
or equivalently, $\lambda_\ideal(z={h_\Lambda}) = \lambda_\res (z={h_\Lambda})$.
The layer $h_\Lambda \leq |z| \leq h_\ideal$ is the so-called ``active layer,''
where MRI operates without affected by ohmic diffusion nor gas stratification. 
The region $|z| \leq h_\Lambda$ is what we call the ``dead zone,'' 
where ohmic diffusion stabilizes the most unstable ideal MRI mode.
For convenience, we regard $h_\Lambda$ as zero when a dead zone is absent.
This is the case for model Ideal.
\item
The third one is $h_\res$ defined by 
\beq
\lambda_\res({z = h_\res})  = h.
\label{eq:h_res}
\eeq
Ohmic diffusion allows the resistive MRI to operate at $h_\res \la |z| \la h_\Lambda$.
At $|z|\la h_\res$, ohmic diffusion stabilizes all the unstable MRI modes.
Note that some previous studies \citep[e.g.,][]{G96,SMUN00} 
used the terminology ``dead zone'' for the region $|z| \leq h_\res$ rather than $|z| \leq h_\Lambda$.
In fact, as we will see in Section~\ref{sec:4},  the set of $h_\Lambda$ and $h_\res$
best characterizes our dead zone.
We regard $h_\res$ as zero when $\lambda_\res$ is less than $h$ at all heights.
This is the case for Y models.
\end{enumerate}

The critical heights in the initial state ($h_{\ideal,0}$, $h_{\Lambda,0}$, and $h_{\res,0}$) 
are shown in Table~1 for all of our 17 simulations.
Using Equations~\eqref{eq:rho} and~\eqref{eq:eta},
one can analytically calculate the initial critical heights for models except FL03L as
\beq
h_{\ideal,0} = \left[2\ln\pfrac{\beta_{z0}}{8\pi^2} \right]^{1/2} h,
\label{eq:z_id0}
\eeq
\beq
h_{\Lambda,0} = \left(\frac{2\ln\Lambda_0^{-1}}{1+(h/h_\eta)^2} \right)^{1/2} h,
\label{eq:z_Lambda0}
\eeq
\beq
h_{\res,0} = 
\Biggl[\dfrac{2\ln\left(8\pi^2\beta_{z0}^{-1}\Lambda_0^{-2}\right)}{1+2(h/h_\eta)^2}\Biggr]^{1/2}h,
\label{eq:z_res0}
\eeq
where 
\beq
\Lambda_0 = \frac{2c_sh}{\eta_\md\beta_{z0}}
\eeq
is the initial Elsasser number at the midplane.

\begin{figure}
\epsscale{.9}
\plotone{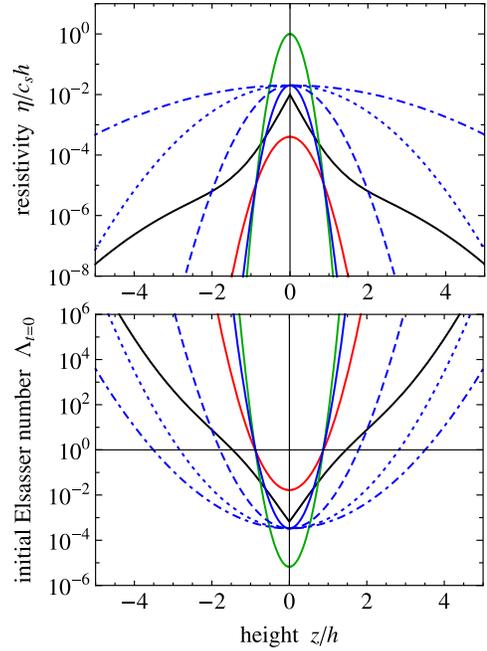}
\caption{Vertical profiles of the ohmic resistivity $\eta$ (upper panel) 
and the initial Elsasser number $\Lambda_{t=0}$ (lower panel) 
for models X0 (blue dot-dashed curve),
X1 (blue dotted curve), X2 (blue dashed curve), X3 (blue solid curve), 
Y3 (red curve), W3 (green curve), and FS03L (black curve).
}
\label{fig:Elsasser}
\end{figure}
Figure~\ref{fig:Elsasser} shows the vertical profiles of
the resistivity $\eta$ and the initial Elsasser number $\Lambda_{t=0}$ 
for some of our models.
The initial midplane Elsasser number $\Lambda_0$ and initial critical heights 
($h_{\ideal,0}, h_{\Lambda,0}, h_{\res,0}$)
 are listed in Table~\ref{tab:param} for all models. 
As one can see from Table~\ref{tab:param} and the lower panel of Figure~\ref{fig:Elsasser}, 
models labeled by the same number are arranged 
so that they have similar values of $h_{\Lambda,0}$.

For turbulent states, we evaluate $v_{Az}$ in the Elsasser number and  
the characteristic wavelengths as $(\ovl{B_z^2}/4\pi\ovl{\rho})^{1/2}$, 
where the overbars denote the horizontal averages.

\section{Simulation Results}\label{sec:4}
\subsection{The Fiducial Model}\label{sec:4.1}
\begin{figure}
\epsscale{1.1}
\plotone{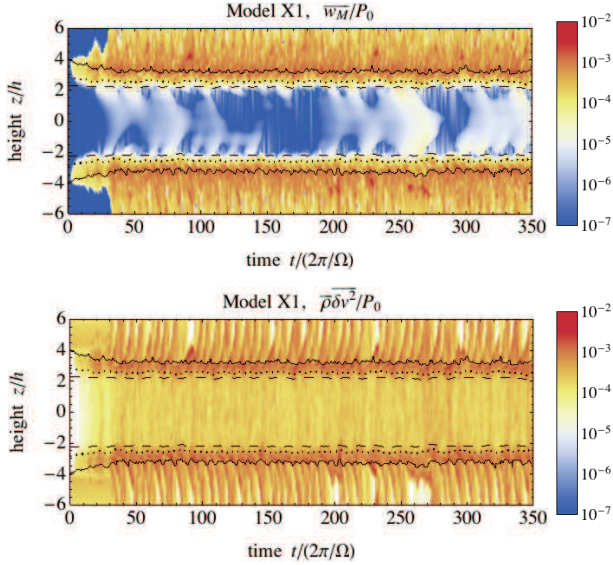}
\caption{Horizontally averaged Maxwell stress $\ovl{w_M}$ (upper panel) 
and density-weighted velocity dispersion $\ovl{\rho}\ovl{\delta v^2}$ (bottom panel)
as a function of time $t$ and height $z$ for model X1.
The solid, dotted, and dashed lines show the critical heights $z=h_\ideal$, $h_\Lambda$, 
and $h_\res$, respectively.
}
\label{fig:tz}
\end{figure}
\begin{figure*}[b]
\epsscale{1.}
\plotone{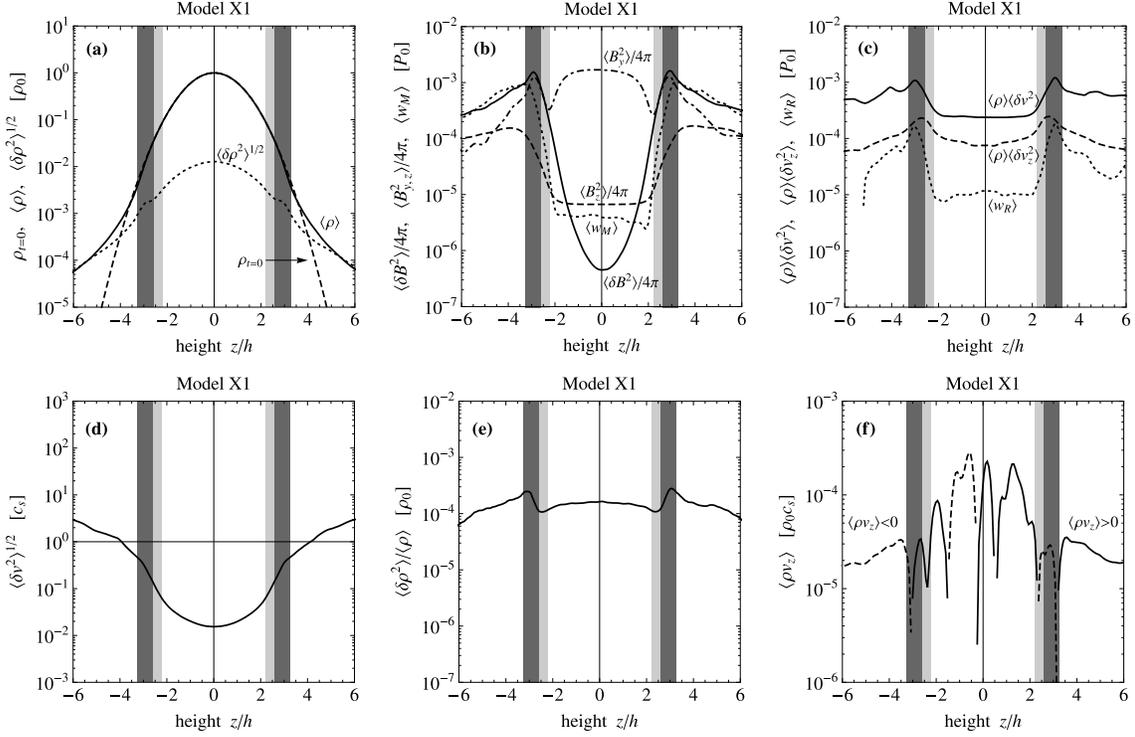}
\caption{Vertical profiles of various quantities averaged over $x$--$y$ 
planes and over a time interval $175~{\rm orbits} < t  < 350~{\rm orbits}$ for model X1.
(a) Initial (dashed curve) and time-averaged (solid curve) gas densities, 
and amplitude of the density fluctuation (dotted curve).
(b) Magnetic energies $\bracket{\delta B^2}/4\pi$ (solid curve), 
$\bracket{B_y^2}/4\pi$ (dot-dashed curve), and
$\bracket{B_z^2}/4\pi$ (dashed curve), 
and Maxwell stress $\bracket{w_M}$ (dotted curve)
normalized by the initial midplane gas pressure $P_0 = \rho_0 c_s^2$.
(c) Density-weighted velocity dispersions $\rhodvtwo$ (thick solid curve) and 
$\rhodvztwo$  (dashed curve), 
and Reynolds stress $\bracket{w_R}$ (dotted curve).
(d) Root-mean-squared random velocity $\bracket{\delta v^2}^{1/2}$.
(e) $\bracket{\delta \rho^2}/\bracket{\rho}$, which represents the thermal energy of fluctuation.
(f) Vertical mass flux $\bracket{\rho v_z}$, plotted  
for positive (solid curve) and negative (dashed curve) values.
The regions shaded in dark and light gray 
indicate where ideal and resistive MRIs operate, respectively (see also Figure~\ref{fig:vert}). 
}
\label{fig:X1_z}
\end{figure*}
We select model X1 as the fiducial model to describe in detail.
Figure~\ref{fig:tz} shows how MRI-driven turbulence reaches a quasi-steady state in run X1.
The upper and lower panels plot the horizontal averages of the Maxwell stress 
$\ovl{w_M} = -\ovl{\delta B_x \delta B_y}/4\pi$ and 
the density-weighted velocity dispersion $\ovl{\rho}\ovl{\delta v^2}$, respectively,
as a function of time $t$ and height $z$.
The solid, dotted, and dashed lines are the loci of the critical heights 
$h_\ideal$, $h_\Lambda$, and $h_\res$, respectively.
As seen in the figure, a quasi-steady state is reached within the first 40 orbits. 
The critical height $h_\ideal$ measured in the quasi-steady state is slightly lower 
than that in the initial nonturbulent state.
This is because  the ideal MRI wavelength $\lambda_\ideal \propto v_{Az}$ is increased by 
the fluctuation in the vertical magnetic field, $\delta B_z^2$. 
By contrast, $h_\Lambda$ and $h_\res$ 
are almost unchanged, because the fluctuation of the magnetic field is suppressed 
in the dead zone.

Figure~\ref{fig:X1_z} shows the vertical structure of the disk averaged over 
a time interval $175 < \Omega t/(2\pi) < 350$.
The dark and light gray bars in each panel 
indicate the heights where ideal and resistive MRIs operate, 
respectively (see also Figure~\ref{fig:vert}). 
The brackets $\bracket{\cdots}$ denote the averages over time and horizontal directions.

In Figure~\ref{fig:X1_z}(a), we compare  the averaged gas density $\bracket{\rho}$
with the initial density given by Equation~\eqref{eq:rho}.
We see that the density is almost unchanged in the disk core ($|z|< h_\ideal$)
but is considerably increased in the atmosphere ($|z|> h_\ideal$).
This is because the magnetic pressure is negligibly small  in the disk core 
but dominates over the gas pressure in the atmosphere.
Figure~\ref{fig:X1_z}(a) also shows the amplitude of the density fluctuation, 
$\bracket{\delta\rho^2}^{1/2}$. 
As one can see, the density fluctuation is small 
($\bracket{\delta\rho^2}^{1/2}\ll \bracket{\rho}$) 
except at $|z|\gg h_\ideal$.

The magnetic activity in the disk can be seen in Figure~\ref{fig:X1_z}(b),
where the vertical profiles of the magnetic energies 
($\bracket{\delta B^2}/4\pi$, $\bracket{B_y^2}/4\pi$, and $\bracket{B_z^2}/4\pi$) 
and Maxwell stress $\bracket{w_M}$ are plotted.
One can see that these quantities peak near the outer boundaries
of the active layers, $|z|\approx h_\ideal$.
This is because the largest channel flows develop  
at locations where $\lambda_\ideal \approx h$ \citep[see, e.g.,][]{SI09}.
In the dead zone, ohmic dissipation suppresses 
the fluctuation in the magnetic fields, $\bracket{\delta B^2}$, 
leaving the initial vertical field ($\bracket{B_z^2} \approx B_{z0}^2$) 
and coherent toroidal fields ($\bracket{B_y^2} \approx \bracket{B_y}^2 $) 
generated by the differential rotation.\footnote{
In our simulations, ohmic resistivity is not high enough to remove shear-generated, 
coherent toroidal fields. In this sense, our dead zone is 
an ``undead zone'' in the terminology of \citet{TS08}.}

\begin{figure*}
\epsscale{0.8}
\plotone{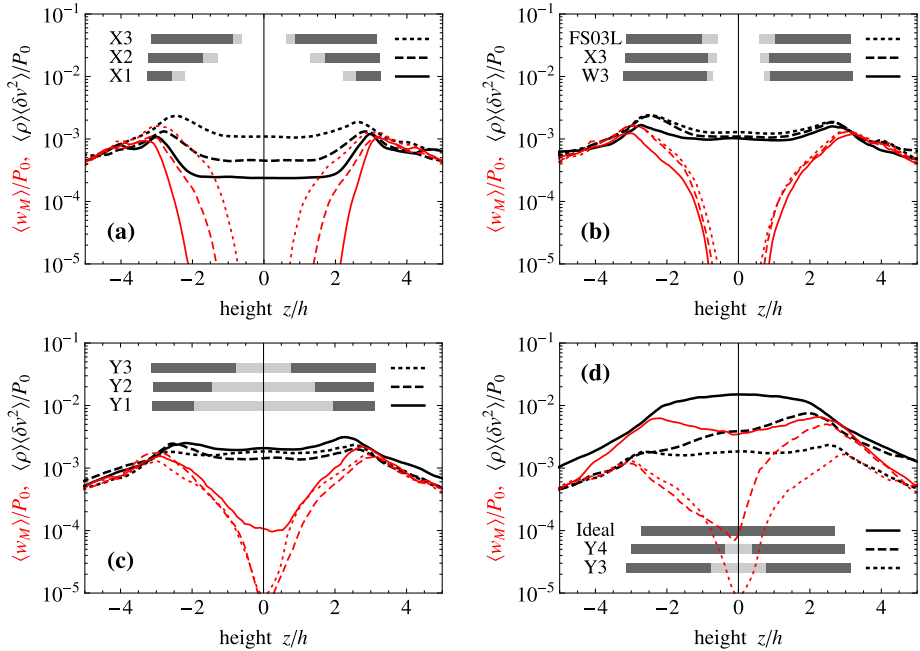}
\caption{
Vertical profiles of temporally and horizontally averaged Maxwell stress 
$\bracket{w_M}$ (red curves) and density-weighted velocity dispersion $\rhodvtwo$ (black curves)
normalized by the initial midplane gas pressure $P_0 = \rho_0 c_s^2$ 
for $\beta_{z0} = 3\times 10^5$ models.
The dark gray bars indicate where MRI operates without affected by ohmic resistivity 
($h_\Lambda < |z| < h_\ideal$), 
while the light gray bars show where MRI operates but is weakened by ohmic resistivity 
($h_\res < |z| < h_\Lambda$).
}
\label{fig:dv2}
\end{figure*}
Figure~\ref{fig:X1_z}(c) shows the density-weighted velocity dispersions 
$\rhodvtwo$ and $\rhodvztwo$  and 
the Reynolds stress $\bracket{w_R} = \bracket{\rho\delta v_x \delta v_y}$.
These quantities characterize the kinetic energy in the random motion of the gas.\footnote{
In the disk core ($|z|\la h_\ideal$),  $\bracket{\rho}\bracket{\delta v^2}$ is approximately equal to   
$\bracket{\rho\delta v^2}$  since the density fluctuation is small (see Figure~\ref{fig:X1_z}(a)).}
Comparing Figures~\ref{fig:X1_z}(b) and (c), 
we find that the drop in these quantities in the disk core
is not as significant as the drop in $\bracket{\delta B^2}$ and $\bracket{w_M}$.
This is an indication that sound waves generated in the active layers 
penetrate deep inside the dead zone \citep{FS03}.
Furthermore, we find that $\rhodvtwo$ is approximately constant, i.e., 
the velocity dispersion $\bracket{\delta v^2}$ 
is inversely proportional to the mean density $\bracket{\rho}$, in the disk core.
This means that the kinetic energy density of fluctuation is nearly constant in the disk core.
This is another indication of sound waves, 
because the amplitude of the velocity fluctuation $\delta v$ 
is generally proportional to $1/\sqrt{\rho}$ for freely propagating sound waves.
The root-mean-squared random velocity $\bracket{\delta v^2}^{1/2}$ 
is shown in Figure~\ref{fig:X1_z}(d).
The random velocity is subsonic in the disk core 
and exceeds the sound speed only in the atmosphere.

An indication of freely sound waves can be also found in the density fluctuation. 
Shown in Figure~\ref{fig:X1_z}(e) is the mean squared density fluctuation 
$\bracket{\delta \rho^2}$ divided by the mean density $\bracket{\rho}$. 
Since $\bracket{\delta\rho^2}^{1/2} \ll \bracket{\rho}$, 
the quantity $\bracket{\delta \rho^2}/\bracket{\rho} \approx \bracket{\delta \rho^2/\rho}$ 
is approximately proportional to  
the thermal energy density of fluctuation, $\bracket{c_s^2\delta \rho^2/2\rho}$.
In the disk core, we see that $\bracket{\delta \rho^2}/\bracket{\rho}$ 
is roughly constant along the vertical direction,
meaning that the amplitude of the density fluctuation, 
$\bracket{\delta \rho^2}^{1/2}$, is proportional to 
the square root of the mean density $\bracket{\rho}$.
The similarity between $\rhodvtwo$ and $\bracket{\delta \rho^2}/\bracket{\rho}$
is peculiar to sound waves, for which  $\delta v/c_s \sim  \delta\rho/\rho$.

Figure~\ref{fig:X1_z}(f) displays the profile of the vertical mass flux $\bracket{\rho v_z}$.
In the atmosphere, the vertical flux is outward, i.e., $\bracket{\rho v_z} >0$ at $z>h_\ideal$
and  $\bracket{\rho v_z} < 0$ at $z<-h_\ideal$.
This outflow results from the breakup of large channel flows 
at the outer boundaries of the active layers, $|z| \approx h_\ideal$ \citep{SI09,SMI10}.
The vertical mass flux reaches a constant value  at $|z|\ga 5h$.
This fact allows us to measure a well-defined outflow flux for each simulation 
(see Section~\ref{sec:5.1.3}).

\subsection{Model Comparison}\label{sec:4.2}
We now investigate how the vertical structure of turbulence depends on 
the resistivity profile and vertical magnetic flux.

Figure~\ref{fig:dv2} displays the temporal and horizontal
averages of the Maxwell stress $\bracket{w_M}$
and the density-weighted velocity dispersion $\bracket{\rho}\bracket{\delta v^2}$ 
as a function of $z$ for various $\beta_{z0} = 3\times 10^5$ models.
As in Figures~\ref{fig:vert} and~\ref{fig:X1_z}, the dark and light gray bars in each panel 
indicate the heights where ideal and resistive MRIs operate, respectively.

The effect of changing the size of the dead zone can be seen in Figure~\ref{fig:dv2}(a),
where models X1, X2, and X3 are compared.
These models are characterized by the same values of $\beta_{z0}$ and $\Lambda_0$ 
but different values of $h_\eta$.
For all the models, $\bracket{w_M}$ sharply falls  at $z \sim h_\res$, 
meaning that $h_\res$ well predicts where the resistivity shuts off the magnetic activity.
By contrast, $\rhodvtwo$ exhibits a flat profile at $|z|\la h_\ideal$ 
with no distinct change across $z=h_\res$ nor $z=h_\Lambda$.
The only clear difference is the value of $\rhodvtwo$ in the disk core, i.e., 
the value is lower when the dead zone is wider.
Note that $\rhodvtwo$ decreases more slowly 
than the column density of the active layers $(h_\Lambda < |z| < h_\ideal)$. 
For example, the active column density in model X1 is 20 times smaller than that in model X3.
However, the midplane value of $\rhodvtwo$ in the former is 
only five times smaller than that in the latter.
This suggests that even a very thin active layer
can provide a large velocity dispersion near the midplane.

Interestingly, the vertical structure of turbulence depends 
on the critical heights ($h_\ideal$, $h_\Lambda$, and $h_\res$)
but are very insensitive to the details of the resistivity profile.
This can be seen in Figure~\ref{fig:dv2}(b), 
where we compare runs with similar critical heights (runs X3, W3, and FS03L).
We see that these models produce very similar vertical profiles of $\bracket{w_M}$ and $\rhodvtwo$
even though they assume quite different resistivity profiles 
(see the upper panel of Figure~\ref{fig:Elsasser}).
This suggests that the vertical structure of turbulence 
is determined by the values of the critical heights.

Importance of distinguishing $h_\Lambda$ and $h_\res$ 
is illustrated in Figures~\ref{fig:dv2}(c) and (d).
These panels compare five models (Y1--4 and Ideal) 
in which the resistive MRI is active at the midplane, i.e., $h_\res = 0$.
Figure~\ref{fig:dv2}(c) shows models with $h_\Lambda > 0.5h$.
We see that the profiles of $\rhodvtwo$ and $\bracket{w_M}$ are
very similar for the three models.
This implies that the vertical structure is determined by the value of $h_\res$
when $h_\Lambda \ga 0.5h$. 
Figure~\ref{fig:dv2}(d) shows what happens when $h_\Lambda$ falls below $0.5h$.
Model Ideal is clearly different from the other models.  
Model Y4 ($h_\Lambda = 0.4$) is interesting because it exhibits {\it both} features.
In the lower half of the disk ($z<0$), the Maxwell stress behaves as 
in the other Y models.
In the upper half ($z>0$), however, the profile of $\bracket{w_M}$ is closer to 
that in model Ideal. 

\begin{figure}
\epsscale{0.8}
\plotone{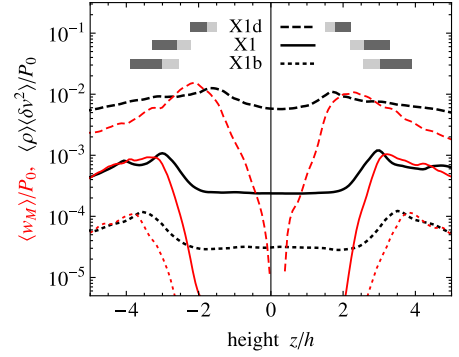}
\caption{
Vertical profiles of temporally and horizontally averaged Maxwell stress $\bracket{w_M}$ (red curves) 
and density-weighted velocity dispersion $\rhodvtwo$ (black curves)
for models with different $\beta_{z0}$. 
The dotted, solid, and dashed curves correspond to models 
X1b, X1, and X1d, respectively.
}
\label{fig:dv2beta}
\end{figure}
Next, we see how the saturation level of turbulence depends on the vertical magnetic flux.
Figure~\ref{fig:dv2beta} shows the vertical profiles of $\bracket{w_M}$ 
and $\bracket{\rho}\bracket{\delta v^2}$ for three runs with different values of $\beta_{z0}$
(X1b, X1, and X1d). 
We see that these values increase with decreasing $\beta_{z0}$.
The peak value of $\bracket{w_M}$ is approximately $10^{-4}P_0$, $10^{-3}P_0$, and $10^{-2}P_0$
for runs X1b, X1, and X1d, respectively ($P_0=\rho_0 c_s^2$ is the initial midplane gas pressure). 
This indicates a linear scaling between the turbulent stress
and $\beta_{z0}^{-1}$.

\section{Scaling Relations and Predictor Functions}\label{sec:5}
Now we seek how the amplitudes of turbulent quantities 
depend on the vertical magnetic flux and the resistivity profile.
We do this in two steps.
First, we derive relations between the amplitudes of turbulent quantities 
and the vertically integrated turbulent stress. 
We then obtain empirical formulae that predict the integrated stress 
as a function of the vertical magnetic flux and the resistivity profile.

\subsection{Scaling Relations between Turbulent 
Quantities and Vertically Integrated Accretion Stresses}\label{sec:5.1}
The ultimate source of the energy of turbulence is the shear motion of the background flow. 
The accretion stress $w_{xy} \equiv w_R+w_M$ determines 
the rate at which the free energy is extracted.
Therefore, we expect that the accretion stress 
is related to the amplitudes of turbulent quantities, 
such as the gas velocity dispersion and outflow mass flux.

To quantify the rate of the energy input in the simulation box,
we introduce the effective $\alpha$ parameter
\beq
\alpha \equiv \frac{\int \bracket{w_{xy}}dz}{\Sigma c_s^2},
\label{eq:alpha}
\eeq 
where $\Sigma = \int \bracket{\rho} dz$ is the gas surface density.
In the classical, one-dimensional viscous disk theory \citep{LP74}, 
the parameter $\alphatot$ is related to the turbulent viscosity $\nu_{\rm trub}$ 
as $\nu_{\rm trub} = (3/2)\alphatot c_s^2/\Omega$, 
where the prefactor $3/2$ comes from the slope of the Keplerian rotation.
Thus, $\alphatot$ also characterizes the vertically integrated mass accretion rate.

As we will see below, it is useful to 
decompose $\alphatot$ as $\alphatot = \alphacore + \alphaatm$, where
\beq
\alphacore  \equiv  \frac{\int_{|z|<h_\ideal}\bracket{w_{xy}}dz}{\Sigma c_s^2}
\label{eq:alphaint}
\eeq
and
\beq
\alphaatm  \equiv  \frac{\int_{|z|>h_\ideal}\bracket{w_{xy}}dz}{\Sigma c_s^2}
\label{eq:alphaext}
\eeq
are the contributions from the disk core ($|z|<h_\ideal$) and 
atmosphere ($|z|>h_\ideal$), respectively.
Table~\ref{tab:avr} shows the values of $\alpha$, $\alphacore$, and $\alphaatm$
as well as the time-averaged critical heights ($h_\ideal$, $h_\Lambda$, $h_\res$) 
for all our simulations. 
\begin{deluxetable*}{lcccllllcll}
\tablecaption{Time-averaged Properties of MHD Simulations}
\tablecolumns{6}
\tablewidth{0pt}
\tablehead{
 \colhead{Model} & 
  \colhead{$\dfrac{h_{\ideal}}{h}$} & 
 \colhead{$\dfrac{h_{\Lambda}}{h}$}  &
  \colhead{$\dfrac{h_{\res}}{h}$} &
 \colhead{$\dfrac{\alphatot}{10^{-3}}$} &
 \colhead{$\dfrac{\alphacore}{10^{-3}}$} &  \colhead{$\dfrac{\alphaatm}{10^{-3}}$} &
 \colhead{$\dfrac{\bracket{\delta v^2}_\md}{10^{-3}c_s^2}$} &
 \colhead{$\dfrac{\bracket{\delta v_z^2}_\md}{\bracket{\delta v^2}_\md}$} &
  \colhead{$\dfrac{\bracket{\delta \rho^2}_\md}{10^{-3}\bracket{\rho}_\md^2}$} &
 \colhead{$\dfrac{\dot{m}_w}{10^{-5}\bracket{\rho}_\md c_s^2}$}
}
\startdata
Ideal & $2.4$ & $0.0$ & $0.0$ & $18$ & $12$ & \phn$6.0$ & $15$ & $0.22$ & \phn$6.4$ & \phn$4.7$ \\ 
X0 & $3.3$ & $3.0$ & $2.8$ & \phn$1.9$ & \phn$0.30$ & \phn$1.6$ & \phn$0.13$ & $0.34$ & \phn$0.12$ & \phn$2.5$ \\ 
X1 & $3.2$ & $2.5$ & $2.2$ & \phn$2.0$ & \phn$0.42$ & \phn$1.5$ & \phn$0.24$ & $0.32$ & \phn$0.16$ & \phn$2.8$ \\ 
X2 & $3.1$ & $1.7$ & $1.3$ & \phn$2.3$ & \phn$0.68$ & \phn$1.6$ & \phn$0.45$ & $0.35$ & \phn$0.27$ & \phn$2.9$ \\ 
X3 & $3.0$ & $0.8$ & $0.6$ & \phn$3.3$ & \phn$1.4$ & \phn$1.9$ & \phn$1.1$ & $0.32$ & \phn$0.66$ & \phn$2.9$ \\ 
Y1 & $2.9$ & $1.6$ & $0.0$ & \phn$4.3$ & \phn$2.0$ & \phn$2.2$ & \phn$2.1$ & $0.26$ & \phn$1.2$ & \phn$3.4$ \\ 
Y2 & $3.0$ & $1.3$ & $0.0$ & \phn$3.6$ & \phn$1.5$ & \phn$2.1$ & \phn$1.4$ & $0.31$ & \phn$0.84$ & \phn$3.4$ \\ 
Y3 & $3.0$ & $0.8$ & $0.0$ & \phn$3.5$ & \phn$1.6$ & \phn$1.9$ & \phn$1.8$ & $0.23$ & \phn$1.2$ & \phn$3.0$ \\ 
Y4 & $2.8$ & $0.4$ & $0.0$ & \phn$7.0$ & \phn$3.8$ & \phn$3.3$ & \phn$3.9$ & $0.26$ & \phn$2.2$ & \phn$4.0$ \\ 
W1 & $3.2$ & $2.6$ & $2.4$ & \phn$1.9$ & \phn$0.37$ & \phn$1.5$ & \phn$0.24$ & $0.36$ & \phn$0.16$ & \phn$3.1$ \\ 
W2 & $3.2$ & $1.7$ & $1.5$ & \phn$2.1$ & \phn$0.59$ & \phn$1.5$ & \phn$0.38$ & $0.37$ & \phn$0.23$ & \phn$2.6$ \\ 
W3 & $3.1$ & $0.9$ & $0.7$ & \phn$2.7$ & \phn$1.1$ & \phn$1.7$ & \phn$1.0$ & $0.39$ & \phn$0.51$ & \phn$2.7$ \\ 
X1a & $4.2$ & $3.2$ & $2.7$ & \phn$0.061$ & \phn$0.014$ & \phn$0.047$ & \phn$0.0097$ & $0.17$ & \phn$0.0083$ & \phn$0.049$ \\ 
X1b & $3.8$ & $2.9$ & $2.5$ & \phn$0.19$ & \phn$0.056$ & \phn$0.13$ & \phn$0.031$ & $0.25$ & \phn$0.023$ & \phn$0.29$ \\ 
X1c & $2.8$ & $2.2$ & $1.9$ & \phn$7.6$ & \phn$1.4$ & \phn$6.2$ & \phn$0.93$ & $0.40$ & \phn$0.50$ & \phn$9.0$ \\ 
X1d & $2.0$ & $1.6$ & $1.4$ & $29$ & \phn$8.0$ & $21$ & \phn$7.0$ & $0.47$ & \phn$1.8$ & $59$ \\ 
FS03L & $3.0$ & $1.0$ & $0.6$ & \phn$3.4$ & \phn$1.5$ & \phn$2.0$ & \phn$1.3$ & $0.35$ & \phn$0.70$ & \phn$3.1$ 
\enddata
\label{tab:avr}
\end{deluxetable*}

\subsubsection{Velocity Dispersion}\label{sec:5.1.1}
\begin{figure}
\epsscale{1.}
\plotone{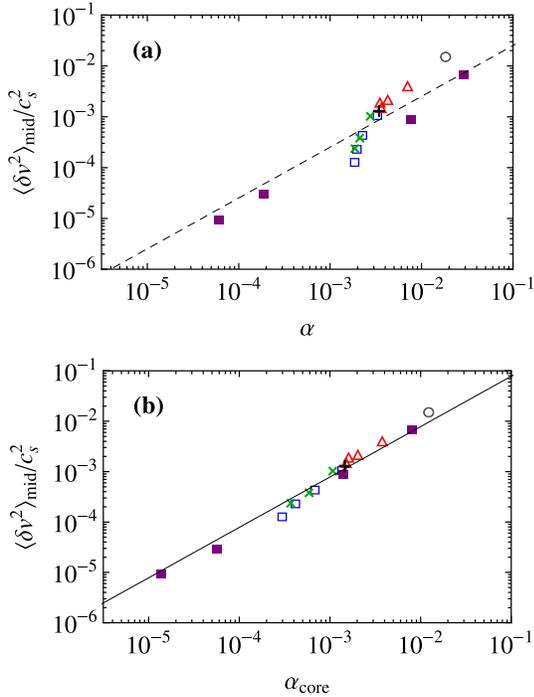}
\caption{Gas velocity dispersion at the midplane, $\bracket{\delta v^2}_{\rm mid}$,
 for all runs presented in this study.
Panels (a) and (b) plot the data versus $\alphatot$ and $\alphacore$, respectively.
The symbols correspond to models Ideal (circle), X0--X3 (open squares), Y1--Y4 (triangles), 
W1--W3 (crosses), X1a--X1d (filled squares), and FS03L (plus sign).
The lines show the best linear fits (Equation~\eqref{eq:dv2mid} for panel (b)).
}
\label{fig:dv2mid}
\end{figure}
Random motion of the gas crucially affects the growth of dust particles 
as it enhances the collision velocity between the particles via friction forces
\citep{W84,J+08}.
Here, we seek how the velocity dispersion $\bracket{\delta v^2}$ is related to 
the integrated accretion stress.

First, we focus on the velocity dispersion at the midplane, $\bracket{\delta v^2}_\md$.
Figure~\ref{fig:dv2mid}(a) shows $\bracket{\delta v^2}_\md$ versus 
the total accretion stress $\alphatot$ for all our runs.
The value of $\bracket{\delta v^2}_\md$ for each run is listed in Table~\ref{tab:avr}.  
One can see a rough linear correlation between the velocity dispersion and the accretion stress
(for reference, a linear fit $\bracket{\delta v^2}_\md =  0.25\alpha c_s^2$ 
is shown by the dashed line). 
However, detailed inspection shows that $\bracket{\delta v^2}_\md$ 
decreases more rapidly than $\alpha$ as the dead zone increases in size.
As found from Table~2, this is because the contribution from 
the atmosphere, $\alphaatm$, is insensitive to the size of the dead zone in the disk core.
In Figure~\ref{fig:dv2mid}(b), we replot the data by replacing $\alpha$ 
with the accretion stress in the disk core, $\alphacore$. 
Comparison between Figures~\ref{fig:dv2mid}(a) and (b) shows that 
$\bracket{\delta v^2}_\md$ more tightly correlates with
$\alphacore$ rather than with $\alpha$.
We find that the data can be well fit by a simple linear relation 
\beq
\bracket{\delta v^2}_\md =  0.78\alphacore c_s^2,
\label{eq:dv2mid}
\eeq
which is shown by the solid line in Figure~\ref{fig:dv2mid}(b).
This result indicates that the accretion stress in the atmosphere does not 
contribute to the velocity fluctuation near the midplane.

\begin{figure}
\epsscale{0.9}
\plotone{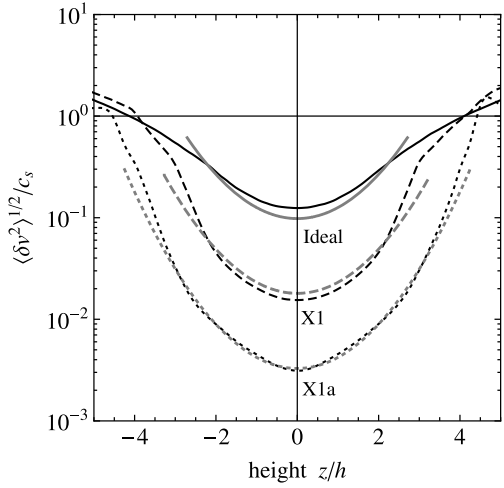}
\caption{Vertical profiles of the random velocity $\bracket{\delta v^2}^{1/2}$ 
directly obtained from MHD simulations (black curves), 
compared with the predictions from Equation~\eqref{eq:dv2}  
with $\alphacore$ taken from Table~\ref{tab:avr} (gray curves).
The solid, dashed, and dotted curves correspond to runs Ideal, X1, and X1a, respectively.
The predicted profiles are plotted only at $|z|<h_\ideal$, where Equation~\eqref{eq:dv2} is valid.
}  
\label{fig:dv}
\end{figure}

Once $\bracket{\delta v^2}_\md$ is known, 
it is also possible to reproduce the vertical profile of the velocity dispersion. 
For the disk core ($|z|< h_\ideal$), we already know 
that  $\bracket{\delta v^2}$ is inversely proportional to the mean gas density $\bracket{\rho}$
and that $\bracket{\rho}$ hardly deviates from the initial Gaussian profile.
From these facts, we can predict the vertical distribution of $\bracket{\delta v^2}$ as
\beqn
\bracket{\delta v^2} &\approx& \bracket{\delta v^2}_\md\frac{\bracket{\rho}_\md}{\bracket{\rho}}
 \approx \bracket{\delta v^2}_\md \exp\pfrac{z^2}{2h^2} \nonumber \\
 &\approx& 0.78\alphacore c_s^2  \exp\pfrac{z^2}{2h^2},
\label{eq:dv2}
\eeqn
where Equation~\eqref{eq:dv2mid} has been used in the final equality.
In Figure~\ref{fig:dv},
we compare the vertical profiles of the random velocity $\bracket{\delta v^2}^{1/2}$ 
directly obtained from runs Ideal, X1, and X1a 
with the predictions from Equation~\eqref{eq:dv2},
where the values of $\alphacore$ are taken from Table~2.
We see that Equation~\eqref{eq:dv2} successfully 
reproduces the vertical profiles of $\bracket{\delta v^2}^{1/2}$ in the disk core.
We remark that Equation~\eqref{eq:dv2} greatly overestimates the velocity dispersion 
at $|z|\gg h_\ideal$, where the gas density can no longer be approximated 
by the initial Gaussian profile (see Figure~\ref{fig:X1_z}(a)). 

\subsubsection{Density Fluctuation}\label{sec:5.1.2}
Density fluctuations generated by MRI-driven turbulence
gravitationally interact with planetesimals and larger solid bodies, 
affecting their collisional and orbital evolution in protoplanetary disks 
\citep{LSA04,NP04,NG10,GNT11}. 
Here, we examine how the amplitude of the density fluctuations is determined 
the vertically integrated accretion stress.

\begin{figure}
\epsscale{1.}
\plotone{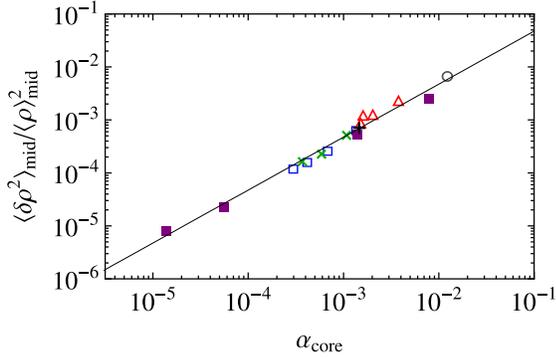}
\caption{
Gas density fluctuation at the midplane, 
$\bracket{\delta \rho^2}_{\rm mid}/\bracket{\rho}_{\rm mid}^2$,
versus $\alphacore$ for all runs presented in this study.
The symbols correspond to models Ideal (circle), X0--X3 (open squares), Y1--Y4 (triangles), 
W1--Y3 (crosses), X1a--X1d (filled squares), and FS03L (plus sign).
The line shows the best linear fit (Equation~\eqref{eq:drho2mid}).  
}
\label{fig:drho2mid}
\end{figure}
As in Section~\ref{sec:5.1.1}, we begin with the analysis of the density fluctuations at the midplane, 
$\bracket{\delta \rho^2}^{1/2}_\md$.
We find from Table~\ref{tab:avr} that $\bracket{\delta \rho^2}_\md$ 
more tightly correlates with $\alphacore$ than with $\alpha$.
Figure~\ref{fig:drho2mid} shows $\bracket{\delta \rho^2}_\md/\bracket{\rho}_\md$
versus $\alphacore$ for all runs. 
The best linear fit is given by
\beq
\bracket{\delta \rho^2}_\md = 0.47 \alphacore \bracket{\rho}_\md^2,
\label{eq:drho2mid}
\eeq
which is shown by the solid line in Figure~\ref{fig:drho2mid}.
If we use this equation with Equation~\eqref{eq:dv2mid},
we can also obtain the relation between the velocity dispersion and density fluctuation,
$\bracket{\delta v^2}_\md/c_s^2 = 1.7\bracket{\delta\rho^2}_\md/\bracket{\rho}^2_\md$.  
This is consistent with the idea
that the fluctuations near the midplane are created by sound waves, 
for which $\delta v/c_s \sim \delta \rho/\rho$ (see also Section~\ref{sec:4.1}).

As shown in Section~\ref{sec:4.1}, $\bracket{\delta \rho^2}$ is roughly proportional to 
$\bracket{\rho}$ along the vertical direction in the disk core.
Hence, if $\bracket{\delta \rho^2}^{1/2}_\md$ is given,
one can reconstruct the vertical profile of the density fluctuation in the disk core
according to 
\beqn
\bracket{\delta \rho^2} &\approx& \bracket{\delta \rho^2}_\md 
\frac{\bracket{\rho}}{\bracket{\rho}_\md} 
\approx \bracket{\delta \rho^2}_\md\exp\left(-\frac{z^2}{2h^2}\right)
\nonumber \\
&\approx& 0.47 \alphacore \bracket{\rho}_\md^2\exp\left(-\frac{z^2}{2h^2}\right),
\label{eq:drho2}
\eeqn
where we have used $\bracket{\rho} \approx \bracket{\rho}_\md \exp(-z^2/2h^2)$
and Equation~\eqref{eq:drho2mid} in the second and third equalities, respectively. 

\subsubsection{Outflow Flux}\label{sec:5.1.3}
We have seen in Section~\ref{sec:4.1} and Figure~\ref{fig:X1_z}(f) that
MRI drives outgoing gas flow at the outer boundaries of the active layers.
The MRI-driven outflow has been first observed by \citet{SI09} in shearing-box simulations
and been recently demonstrated by \citet{Flock+11} in global simulations. 
\citet{SI09} and \cite{SMI10} point out that
this outflow might contribute to the dispersal of protoplanetary disks, 
although it is still unclear whether the outflow can really escape from the disks (see below). 
Meanwhile, MRI also contributes to the accretion of the gas in the radial direction.
For consistent modeling of these two effects,
we seek how the accretion stress and outflow flux are correlated with each other.

We evaluate the outflow mass flux in the following way.
As seen in Section~\ref{sec:4.1}, the temporally and horizontally averaged vertical mass flux 
$\bracket{\rho v_z}$ is nearly constant at heights $|z| \ga 5h$. 
Using this fact, we define the outflow mass flux $\dot{m}_w$ 
as the sum of $|\bracket{\rho v_z}|$ averaged near upper and lower boundaries,
\beq
\dot{m}_w = \frac{\int_{h_{\rm bnd}-h}^{h_{\rm bnd}}|\bracket{\rho v_z}|dz}{h} 
+ \frac{\int_{-h_{\rm bnd}}^{-h_{\rm bnd}+h}|\bracket{\rho v_z}|dz}{h},
\eeq
where $h_{\rm bnd} = 5\sqrt{2}h \approx 7h$ 
is the height of the upper and lower boundaries of the simulation box.

\begin{figure}
\epsscale{1.}
\plotone{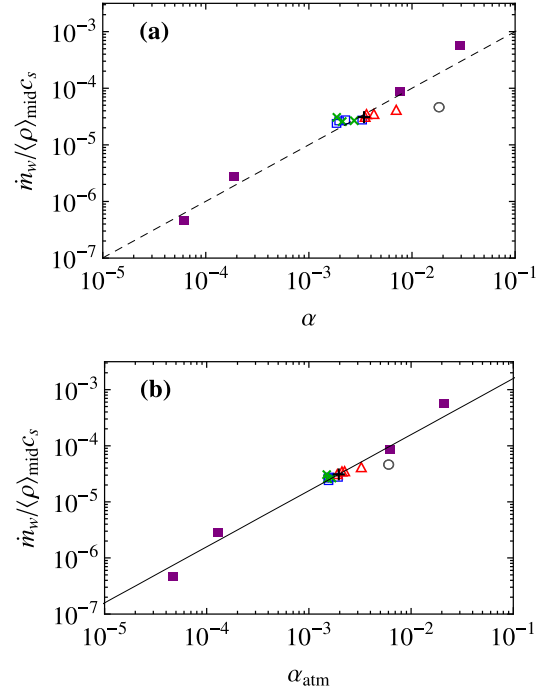}
\caption{Outflow mass flux $\dot{m}_w$ normalized by $\bracket{\rho}_\md c_s$
for all runs presented in this study.
Panels (a) and (b) plot the data versus $\alphatot$ and $\alphaatm$, respectively.
The symbols correspond to models Ideal (circle), X0--X3 (open squares), Y1--Y4 (triangles), 
W1--W3 (crosses), X1a--X1d (filled squares), and FS03L (plus sign).
The lines show the best linear fits (Equation~\eqref{eq:wind} for panel (b)).
}
\label{fig:wind}
\end{figure}
Figure~\ref{fig:wind}(a) shows $\dot{m}_w$ normalized by $\bracket{\rho}_\md c_s$ versus
$\alphatot$ for all simulations. 
The dimensionless quantity $\dot{m}_w/(\bracket{\rho}_\md c_s)$ 
is equivalent to $C_w$ used in \citet{SMI10}.
For model Ideal, the value 
$\dot{m}_w = 5\times 10^{-5}\bracket{\rho}_\md c_s$ is consistent 
with the $\beta_{z0} = 10^5$ ideal run of \cite{SI09}.
The dashed line shows the best linear fit 
$\dot{m}_w/(\bracket{\rho}_\md c_s) = 0.015\alphatot$.
It can be seen that the linear fit captures a rough trend
but still considerably overestimates the outflow flux for models Ideal and Y4.
As seen in Table~\ref{tab:avr}, these are the models in which $\alphacore$ dominates over $\alphaatm$.
This implies that the turbulence in the disk core  (which is the source of $\alphacore$)
does not contribute to the outflow.
In Figure~\ref{fig:wind}(b), we replot the data by replacing $\alphatot$ with $\alphaatm$.
We find that $\dot{m}_w$ more tightly correlates with $\alphaatm$ than with $\alphatot$.
The best linear fit is found to be 
\beq
\dot{m}_w = 0.016\alphaatm \bracket{\rho}_\md c_s.
\label{eq:wind}
\eeq
This result is consistent with the idea that the outflow 
is driven at the outer boundaries of the active layers \citep{SI09}, 
because the dominant contribution to $\alphaatm$ comes from heights very close to $h_\ideal$.

Although outflow from the simulation box is a general phenomenon in our simulations, 
it is unclear whether the outflow leaves or returns to the disk.
In fact, the outflow velocity observed in our simulations does not exceed the sound speed 
even at the vertical boundaries. Since the escape velocity is higher than the sound speed,
this means that the outflow does not have an outward velocity enough to escape out of the disk.
Acceleration of the outflow beyond the escape velocity has not been directly demonstrated by
previous simulations as well \citep{SMI10,Flock+11}.
However, \citet{SMI10} point out a possibility that magnetocentrifugal 
forces and/or stellar winds could accelerate the outflow to the escape velocity. 
If the escape of the outflow will be confirmed in the future, 
our scaling formula for $\dot{m}_w$ will certainly become 
a useful tool to discuss the dispersal of protoplanetary disks.

\subsection{Saturation Predictors for the Accretion Stresses}\label{sec:5.2}
In the previous subsection, we have shown that the amplitudes of various turbulent quantities 
scale with the vertically integrated stresses $\alphacore$ and $\alphaatm$.
The next step is to find out how to predict $\alphacore$ and $\alphaatm$ 
in the saturated state from the vertical magnetic flux $B_{z0}$ (or equivalently $\beta_{z0}$)
and the resistivity profile $\eta$.
As shown in Section~\ref{sec:4.2}, the turbulent state of a disk depends on the resistivity 
only through the critical heights of the dead zone, $h_\Lambda$ and $h_\res$.
Furthermore,  
the values of $h_\Lambda$ and $h_\res$ are only weakly affected by
the nonlinear evolution of MRI 
since the fluctuations in $B_z$ and $\rho$ are small inside the dead zone (see Section~\ref{sec:4.1}).
Therefore, we expect that the effect of the resistivity can be well predicted by
the values of $h_\Lambda$ and $h_\res$ in the initial state, i.e.,  $h_{\Lambda,0}$ and $h_{\res,0}$.
With this expectation, we try to derive saturation predictors for
$\alphacore$ and $\alphaatm$ as a function of $\beta_{z0}$,  $h_{\Lambda,0}$, and $h_{\res,0}$.

\begin{figure}
\epsscale{1.}
\plotone{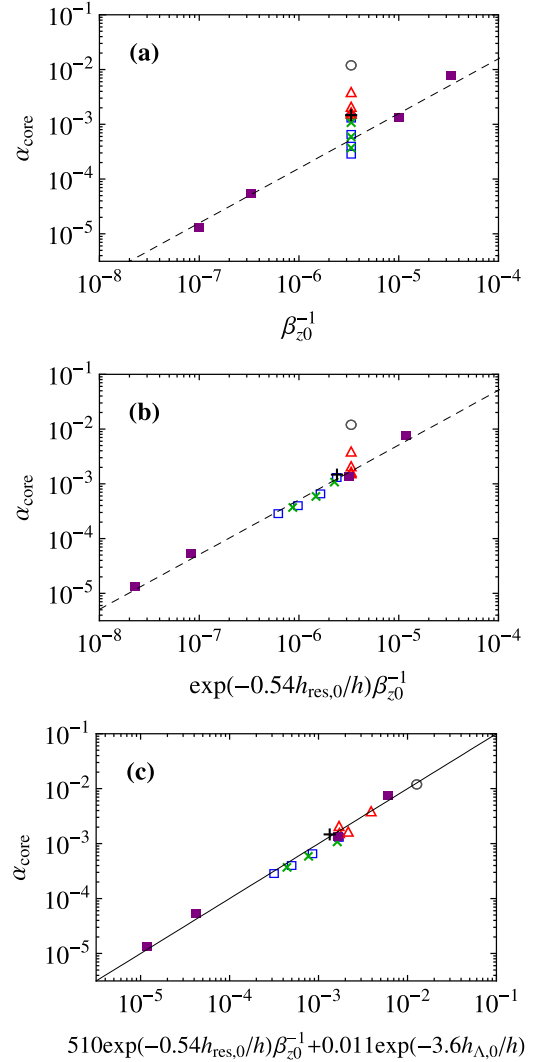}
\caption{Disk core accretion stress $\alphacore$ for all our simulations.
(a) Vs.~the inverse initial plasma beta $\beta_{z0}^{-1}$.
(b) Vs.~$\beta_{z0}^{-1}$ multiplied by $\exp(-0.54h_{\res,0}/h)$
(see also Figure~\ref{fig:alphahres}).
(c) Vs.~the final predictor function, Equation~\eqref{eq:pred_core} (solid line).
The symbols correspond to models Ideal (circle), X0--X3 (open squares), Y1--Y4 (triangles), 
W1--Y3 (crosses), X1a--X1d (filled squares), and FS03L (plus sign).
The dashed lines in panels (a) and (b) are linear fits, shown only for reference.
}
\label{fig:alphaint}
\end{figure}
\begin{figure}
\epsscale{1.}
\plotone{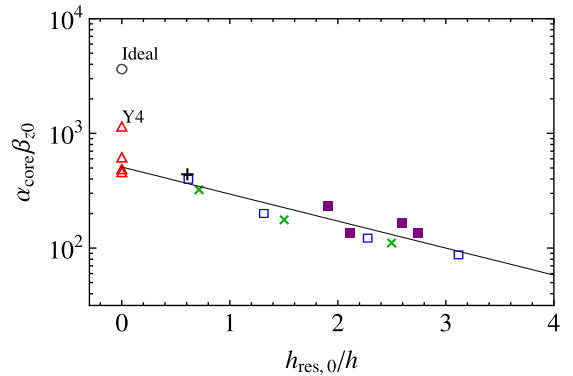}
\caption{$\alphacore\beta_{z0}$ versus $h_{\res,0}$ for all runs.
The symbols correspond to models Ideal (circle), X0--X3 (open squares), Y1--Y4 (triangles), 
W1--W3 (crosses), X1a--X1d (filled squares), and FS03L (plus sign).
The solid line shows an exponential fit for runs except Ideal and Y4 
(Equation~\eqref{eq:pred_core_hres}).
}
\label{fig:alphahres}
\end{figure}
First, we focus on $\alphacore$.
Figure~\ref{fig:alphaint}(a) plots $\alphacore$ versus $\beta^{-1}_{z0}$ for all our simulations.
We see that $\alphacore$ scales roughly linearly with $\beta^{-1}_{z0}$.
The deviation from the linear scaling is expected to come from 
the difference in the dead zone size, i.e., $h_{\Lambda,0}$ and $h_{\res,0}$.
In Figure~\ref{fig:alphahres}, 
we plot the product $\alphacore\beta_{z0}$ as a function of $h_{\res,0}$. 
For models except Ideal and Y4,  
we find that $\alphacore\beta_{z0}$ is well predicted by a simple formula 
\beq
\alphacore\beta_{z0} = 510 \exp(-0.54h_{\res,0}/h).
\label{eq:pred_core_hres}
\eeq
Figure~\ref{fig:alphaint}(b) replot the data in Figure~\ref{fig:alphaint}(a) 
by replacing $\beta^{-1}_{z0}$ with $\exp(-0.54h_{\res,0}/h)\beta^{-1}_{z0}$.
For models Ideal and Y4, Equation~\eqref{eq:pred_core_hres} underestimates $\alphacore$.
As explained in Section~\ref{sec:4.2}, 
these models exhibit higher magnetic activity near the midplane than the other models
because of no or a thin dead zone $(2h_\Lambda < h)$.
We expect that the higher magnetic activity 
gives additional contribution to $\alphacore$.
Taking into account this effect, we arrive at the final predictor function,
\beq
\alphacore = 510\exp(-0.54h_{\res,0}/h)\beta^{-1}_{z0} + 0.011\exp(-3.6h_{\Lambda,0}/h).
\label{eq:pred_core}
\eeq
Here, the numerical factors $0.011$ and $3.6$ appearing in the second term have been chosen to 
reproduce the results of runs Ideal and Y4, respectively.
Figure~\ref{fig:alphaint}(c) compares the final fitting formula with the numerical data.
It can be seen that Equation~\eqref{eq:pred_core} well predicts $\alphacore$ for all our models.
Note that the second term of the predictor function is assumed 
to have no explicit linear dependence on $\beta_{z0}^{-1}$ unlike the first term. 
In fact, it is possible to reproduce our data by multiplying the second term by a prefactor 
$3\times 10^5/\beta_{z0}$.
However, as we will see below, the absence of the prefactor 
makes the predictor function consistent with the results of ideal MHD simulations in the literature.

\begin{figure}
\epsscale{1.}
\plotone{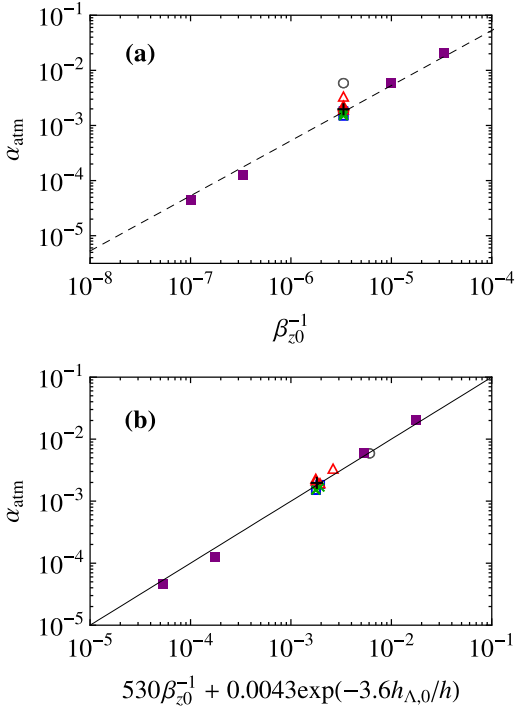}
\caption{
Atmosphere accretion stress $\alphaatm$ for all our simulations.
(a) Vs. the inverse initial plasma beta $\beta_{z0}^{-1}$.
(b) Vs. the final predictor function, Equation~\eqref{eq:pred_atm} (solid line).
The symbols correspond to models Ideal (circle), X0--X3 (open squares), Y1--Y4 (triangles), 
W1--W3 (crosses), X1a--X1d (filled squares), and FS03L (plus sign).
The dashed line in panel (a) is a linear fit, only shown for reference.
}
\label{fig:alphaext}
\end{figure}
The predictor function for $\alphaatm$ can be obtained in a similar way.
Figure~\ref{fig:alphaext}(a) shows $\alphaatm$ versus $\beta_{z0}^{-1}$ for all our runs. 
We find that a simple linear relation $\alphaatm = 530\beta_{z0}^{-1}$ well fits to the data 
except for models Ideal and Y4.
This means that $\alphaatm$ is characterized only by $\beta_{z0}^{-1}$ 
as long as the dead zone is thick $(2h_\Lambda > h)$.
To take into account the cases of thin dead zones, 
we add a term proportional to $\exp(-3.6h_{\Lambda,0}/h)$ 
as has been done for $\alphacore$, and obtain
\beq
\alphaatm = 530\beta^{-1}_{z0} +  0.0043\exp(-3.6h_{\Lambda,0}/h),
\label{eq:pred_atm}
\eeq
where the prefactor $0.0043$ for the second term has been determined 
to fit to the result of run Ideal.
As seen in Figure~\ref{fig:alphaext}(b), 
Equation~\eqref{eq:pred_atm} well predicts the value of $\alphaatm$ for all our runs.

\begin{figure}
\epsscale{0.9}
\plotone{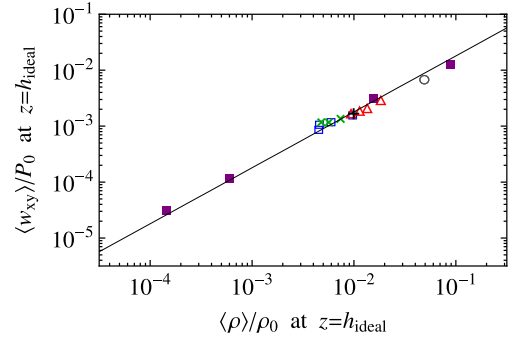}
\caption{Accretion stress $\bracket{w_{xy}}$ versus 
gas density $\bracket{\rho}$ at the upper boundary of the active layer ($z=h_\ideal$) for all runs. 
The symbols correspond to models Ideal (circle), X0--X3 (open squares), Y1--Y4 (triangles), 
W1--Y3 (crosses), X1a--X1d (filled squares), and FS03L (plus sign).
The solid line shows a linear fit (Equation~\eqref{eq:w_rho}).
}
\label{fig:wrho}
\end{figure}
Our predictor functions indicate that the vertically integrated accretion stress is inversely 
proportional to $\beta_{z0}$ when a large dead zone is present ($h_\Lambda \ga h$).
As we show below, this dependence originates from the magnitude of the accretion stress  
at the outer boundaries of the active layers, $|z|\approx h_\ideal$.
When a dead zone exists,  the dominant contribution to $\alpha$ comes from 
the accretion stress at that location (see Figures~\ref{fig:dv2} and \ref{fig:dv2beta}).
As shown Figure~\ref{fig:wrho}, our simulations suggest that
the accretion stress at $|z|= h_\ideal$ obeys a simple relation 
\beq
\bracket{w_{xy}}(h_\ideal) \approx 0.18 \bracket{\rho}(h_\ideal) c_s^2.
\label{eq:w_rho}
\eeq
This means that the averaged accretion stress at $|z|=h_\ideal$
is $18\%$ of the averaged gas pressure $\bracket{\rho c_s^2}$ at the same height.
By the definition of $h_\ideal$, the gas density at $|z|=h_\ideal$ 
is related to $\bracket{B_z^2}$ at the same height
 as $\bracket{\rho}(h_\ideal) = (2\pi)^2  \bracket{B_z^2}(h_\ideal)/4\pi c_s^2$.
Since our simulations suggest $\bracket{B_z^2}(h_\ideal) \sim 10 B_{z0}^2$,
the relation means  
$\bracket{\rho}(h_\ideal) \sim 10(2\pi)^2 B_{z0}^2/4\pi c_s^2 \sim 10^3 \beta_{z0}^{-1}\rho_0$.  
Using this fact, Equation~\eqref{eq:w_rho} can be rewritten into the linear relation 
between $\bracket{w_{xy}}(h_\ideal)$ and $\beta_{z0}^{-1}$:
\beq
\bracket{w_{xy}}(h_\ideal) \sim 100 \beta_{z0}^{-1}\rho_0c_s^2.
\eeq
When a dead zone is present, 
the level of $\alpha$ is determined by $\bracket{w_{xy}}(h_\ideal)$ (see above),
so we have $\alpha \propto \beta_{z0}^{-1}$.

\begin{figure}
\epsscale{0.9}
\plotone{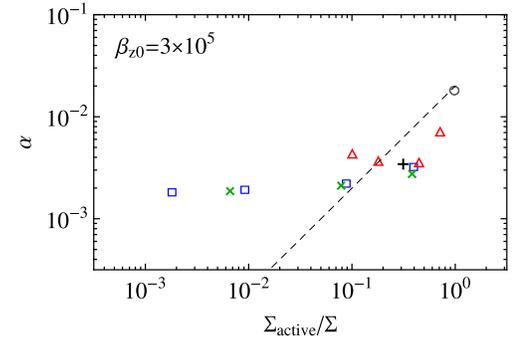}
\caption{Vertically integrated accretion stress $\alpha$ versus 
the column density $\Sigma_{\rm active}$ of the active layer 
for $\beta_{z0} = 3 \times 10^5$ models. 
The symbols correspond to models Ideal (circle), X0--X3 (open squares), Y1--Y4 (triangles), 
W1--Y3 (crosses), and FS03L (plus sign).
The dashed line is a linear function, only shown for reference.
}
\label{fig:alphaSigma}
\end{figure}
We remark that the vertically integrated stress 
does not scale linearly with the column density of active layers.  
This is shown in Figure~\ref{fig:alphaSigma}, 
where we compare $\alpha$ with the column density $\Sigma_{\rm active}$
of the active region $h_\Lambda < |z| < h_\ideal$.
We see that $\alpha$ decreases much more slowly than $\Sigma_{\rm active}$
when $\Sigma_{\rm active}$ is less than $10\%$ of the total gas surface density.
This reflects the fact that the dominant contribution to $\alpha$ comes from the outer boundaries 
of the active zones, $|z|\approx h_\ideal$.

It is useful to see how the predictor functions work when a dead zone is absent.
If  $h_{\Lambda,0} = h_{\res,0} = 0$, 
Equations~\eqref{eq:pred_core} and \eqref{eq:pred_atm} predict the total accretion stress 
$\alpha = 1.0\times 10^3\beta_{z0}^{-1} + 0.015$.
This implies that $\alpha$ is constant ($\alpha \approx 10^{-2}$) 
for $\beta_{z0} \ga 10^5$ and increases linearly with $\beta_{z0}^{-1}$ 
($\alpha \approx 10^{-2}(10^5/\beta_{z0})$) for $\beta_{z0} \la 10^5$.
Strikingly, this prediction is consistent with the finding 
by \citet[][see their Figure 2]{SMI10}.
The existence of the floor value $\alpha \approx 10^{-2}$ 
at low net vertical magnetic fluxes (i.e., at high $\beta_{z0}$)
 is also supported by recent stratified MHD simulations 
 with zero net flux \citep{DSP10}.
These facts suggest that our predictor functions are applicable 
even when a dead zone is absent.

\section{Vertical Diffusion Coefficient}\label{sec:6}
As seen in Section~\ref{sec:4}, sound waves excited in the upper layers 
create fluctuations in the gas velocity near the midplane. 
It has been well known that fully developed MRI-driven turbulence 
causes the diffusion of small dust particles \citep{JK05,TCS10}.
However, it has not been fully understood how the sound waves propagating 
inside a dead zone affect the dynamics of dust particles there.
For example, \cite{SI09} speculated that the sound waves might promote 
dust sedimentation by transferring the downward momentum to dust particles.
On the other hand, \citet{TCS10} reported that the waves excite vertical oscillation of 
dust particles deep inside the dead zone and thus prevent the formation of a thin dust layer.
Since dust sedimentation is crucial to planetesimal formation via gravitational instability, 
it is worth addressing here how it is affected by the velocity dispersion created by sound waves. 
\begin{figure*}
\epsscale{1.0}
\plotone{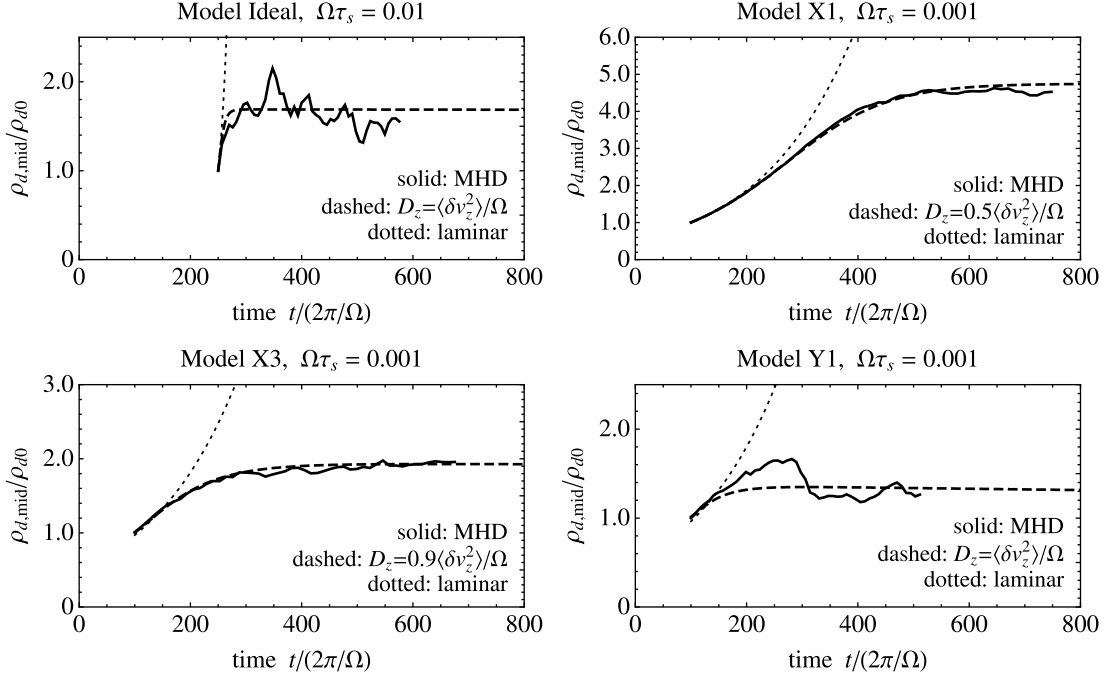}
\caption{
Temporal evolution of the dust density at the midplane in various MHD runs.
The four panels show the results for models Ideal (upper left), X1 (upper right), X3 (lower left),
and Y1 (lower right).
The solid curves show the horizontally averaged dust density $\rho_{d,\md}$ observed in the MHD runs,
while the dotted curve show the evolution of $\rho_{d,\md}$ in a laminar disk.
The dashed curves are the predictions from the one-dimensional advection-diffusion
equation~(Equation~\eqref{eq:advdiff}).
The diffusion coefficient $D_z$ in Equation~\eqref{eq:advdiff} is assumed to be proportional to
$\bracket{\delta v_z^2}/\Omega$, and the proportionality factor has been chosen to 
best reproduce the evolution of $\rho_{d,\md}$ in the MHD runs.
}
\label{fig:rhodmid}
\end{figure*}

Here, we focus on the dynamics of small dust particles, 
and model the swarm of the particles as a passive scalar 
as was previously done by \citet{JK05} and \citet{TCS10}.
We assume that dust particles are so small and their stopping time $\tau_s$ 
is much shorter than the turnover time of turbulence ($\sim \Omega^{-1}$).
We also assume that the dust density is lower than the gas density 
and hence the dust has no effect on the gas motion. 
Under these assumptions, the velocity of dust particles relative to the gas 
can be approximated by the terminal velocity 
${\bf V}_T = - \Omega^2 \tau_s z \hat{{\bf z}}$,
where $\hat{{\bf z}}$ is the unit vector for the $z$--direction.  
Then, the equation of continuity for dust is given by 
\beq
\frac{\pd \rho_d}{\pd t} + \nabla\cdot[\rho_d({\bf v} + {\bf V}_T)] = 0,
\label{eq:eoc_dust}
\eeq
where $\rho_d$ is the dust density. 
Equation~\eqref{eq:eoc_dust} has an advantage 
that the time step can be taken longer than $\tau_s$ in numerical calculation.

We have solved Equation~\eqref{eq:eoc_dust} for four models (Ideal, X1, X3, and Y1)
with the initial condition that the dust-to-gas mass ratio $f \equiv \rho_d/\rho$ 
is constant throughout the simulation box.
To extract the effect of the quasi-stationary turbulence,
we insert the dust 100 (for models X1, X3, and Y1) or 250 (for model Ideal) 
orbits after the beginning of the MHD calculations.
The stopping time $\tau_s$ is set to $\tau_s = 0.01\Omega^{-1}$ 
for model Ideal and $\tau_s = 0.001\Omega^{-1}$ for the other models.
The longer $\tau_s$ has been adopted for model Ideal 
to allow the dust to settle appreciably in the stronger turbulence.
In reality, the stopping time of a dust particle 
depends on the gas density and hence on $z$, 
but we ignore this dependency for simplicity.

Figure~\ref{fig:rhodmid} shows the temporal evolution of the dust density at the midplane, 
$\rho_{d,\md}$, for the four MHD runs.
The solid curves show the horizontally averaged $\rho_{d,\md}$ observed in the MHD runs.
For comparison, the evolution of $\rho_{d,\md}$ in a hydrostatic, laminar disk is also shown
by the dotted curves.
Irrespectively of the presence or absence of a dead zone, 
the dust density observed in the MHD runs 
is higher than that in a laminar disk at all moments.
This means that sound waves propagating in a dead zone 
do not promote but prevent dust settling as turbulence does in an active zone. 

To illustrate more clearly the diffusive nature of the velocity dispersion in a dead zone, 
we try to compare the above results with a simple advection-diffusion theory.
Here, we consider a one-dimensional advection-diffusion equation \citep{DMS95}
\beq
\frac{\pd \rho_d}{\pd t} = -\frac{\pd(\rho_dV_T)}{\pd z}
+ \frac{\pd}{\pd z}\left( D_{z}\rho \frac{\pd}{\pd z}\frac{\rho_d}{\rho} \right),
\label{eq:advdiff}
\eeq
where $V_T$ is the terminal velocity given above and 
$D_{z}$ is the vertical diffusion coefficient for dust.
If the dust particles are sufficiently small ($\tau_s\Omega \ll 1$), 
$D_{z}$ is equal to the diffusion coefficient for gaseous contaminants \citep[e.g.,][]{YL07}.
The first term in the right-hand side of  Equation~\eqref{eq:advdiff} 
represents the downward advection of dust due to settling, 
while the second term represents the diffusion of dust in the stratified gas. 
For disks with no or a small ($|z|\la h$) dead zone, 
it is known that Equation~\eqref{eq:advdiff} well describes the evolution of $\rho_d$ 
if the diffusion coefficient is assumed to be \citep{FP06}
\beq
D_z \approx \bracket{\delta v_z^2}/\Omega.
\label{eq:Dzz}
\eeq

\begin{figure}
\plotone{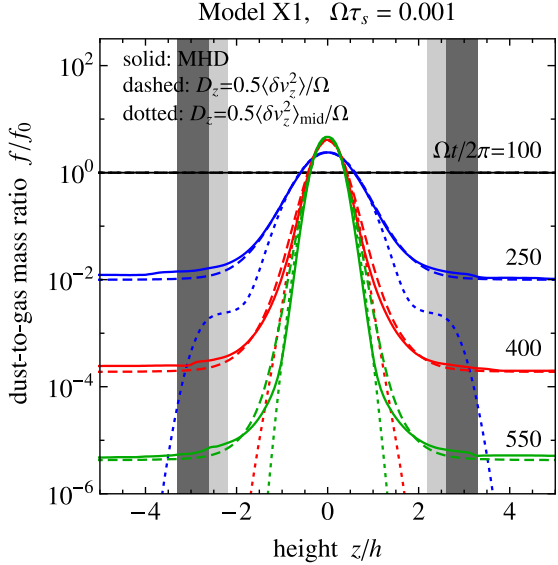}
\caption{Snapshots of the vertical distribution of the dust-to-gas mass ratio $f = \rho_d/\rho$
at $t = 100$, 250, 400, and 550 orbits for model X1.
The solid curves show the horizontally averaged $f$ observed 
in the MHD simulation.
The dashed curves show the solutions to the one-dimensional advection-diffusion
equation~(Equation~\eqref{eq:advdiff}) with $D_z(z) = 0.5\bracket{\delta v_z^2}(z)/\Omega$.
The dotted curves show the solution to Equation~\eqref{eq:advdiff} 
with a constant diffusion coefficient $D_z = 0.5\bracket{\delta v_z^2}_{\rm mid}/\Omega$.
}
\label{fig:fdg_X1}
\end{figure}
We here examine whether Equations~\eqref{eq:advdiff} and \eqref{eq:Dzz} work well
even when a dead zone is present.
We solve Equation~\eqref{eq:advdiff} with $D_z = b\bracket{\delta v_z^2}/\Omega$,
where the vertical distribution of $\bracket{\delta v_z^2}$ is taken from 
temporally and horizontally averaged MHD data and $b$ is 
a dimensionless fitting parameter.
The dashed curves in Figure~\ref{fig:rhodmid} show the predictions 
by the advection-diffusion model, where $b$ is set to be $1.0$, $0.5$, $0.9$, and $1.0$
for runs Ideal, X1, X3, and Y1, respectively.
It can be seen that the advection-diffusion model with $b\sim 1$ successfully reproduces 
the long-term evolution of the observed $\rho_{d,\md}$ for all the models.
It is striking that a constant $b$ well reproduces the evolution of the dust density 
at {\it all} heights, as is shown in Figure~\ref{fig:fdg_X1}. 
In this figure, the solid and dashed curves show the vertical distribution 
of the dust-to-gas mass ratio $f = \rho_d/\rho_g$ observed in run X1 and predicted by  
the advection-diffusion model with $b=0.5$, respectively.
In this run, the boundaries between the active and dead zones are located 
at $|z| = h_\Lambda \approx 2.5h$ (see Table~\ref{tab:avr}).
However, Equation~\eqref{eq:advdiff} successfully predicts the evolution of $f$ 
even if we do not change the value of $b$ across the boundaries.
This fact supports the idea that sound waves propagating across a dead zone contribute 
to the diffusion of dust particles just as turbulence does in active zones.

It is worth mentioning here that the diffusion coefficient 
$D_z$ increases with $|z|$ as has been pointed out by \citet{T+06} and \citet{FN09}.
This effect is particularly significant at high altitudes where 
the gas density is much lower than that at the midplane,
because $D_z \propto \bracket{\delta v^2_z}$ 
is roughly proportional to the inverse of the gas density.\footnote{
\citet{FN09} used a diffusion coefficient quadratic in $z$ 
to explain the dust distribution in their MHD simulation.
Our finding $D_z \propto \rho^{-1}$ does not contradict their assumption 
because $\rho^{-1} \propto \exp(z^2/2h^2) \approx 1 + z^2/2h^2$ 
near the midplane.}
The dotted curves in Figure~\eqref{fig:fdg_X1} show 
how Equation~\eqref{eq:advdiff} would fail to predict dust evolution 
if one assumed a constant diffusion coefficient 
$D_z = 0.5\bracket{\delta v^2_z}_{\rm mid}/\Omega$.
We see that the constant diffusion coefficient model significantly 
underestimates the dust density at $|z| \gg h$.
This fact will merit consideration when modeling 
the chemical evolution of protoplanetary disks, 
in which the vertical mixing of molecules is of importance \citep{HNWM11}.

Finally, we give a simple analytic recipe for the vertical distribution of $D_z$.
It is useful to rewrite Equation~\eqref{eq:Dzz} in terms of $\bracket{\delta v^2}$,
for which the scaling relation (Equation~\eqref{eq:dv2}) and predictor function 
(Equation~\eqref{eq:pred_core}) are available.
Table~2 lists the ratio of $\bracket{\delta v_z^2}_\md$ to $\bracket{\delta v^2}_\md$
for all our simulations.
It can be seen that 
$\bracket{\delta v_z^2}_\md \approx (32\% \pm15\% )\times \bracket{\delta v^2}_\md$,
indicating that $\bracket{\delta v_z^2}_\md$ is roughly equal to
 a third of $\bracket{\delta v^2}_\md$.
Furthermore, the ratio $\bracket{\delta v_z^2}/\bracket{\delta v^2}$ 
is approximately constant in the disk core, as is illustrated in Figure~\ref{fig:X1_z}(c).
Based on these facts, we approximate $\bracket{\delta v_z^2}$ as $\bracket{\delta v^2}/3$
in the disk core.
Using this approximation together with 
the scaling relation for $\bracket{\delta v^2}$ (Equation~\eqref{eq:dv2}),
we rewrite Equation~\eqref{eq:Dzz} as 
\beqn
D_z &\approx& \frac{1}{3}\bracket{\delta v^2}/\Omega \nonumber \\
&\approx& 0.3(\alphacore c_s^2/\Omega)\exp\pfrac{z^2}{2h^2}.
\label{eq:Dz}
\eeqn
If one uses this equation together with the predictor function for $\alphacore$ 
(Equation~\eqref{eq:pred_core}), one can calculate the vertical distribution of $D_z$
in the disk core for given $\beta_{z0}$ and $\eta$.

\section{Discussion: Effects of Numerical Resolution}\label{sec:7}
All MHD simulations presented in the previous sections were performed 
with the numerical resolution of $40\times 80\times 200$ grid cells 
for the simulation box of size $2\sqrt{2}h\times 8\sqrt{2}h \times 10\sqrt{2}h$.  
Here, we examine how the numerical resolution affects the saturated state of turbulence.

\begin{figure}
\epsscale{0.9}
\plotone{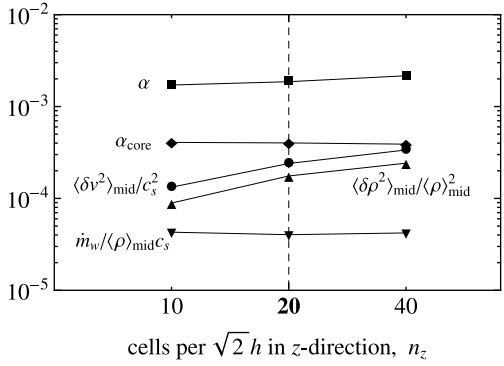}
\caption{
Saturated values of various quantities versus numerical resolution for model X1.
From top to bottom: $\alpha$, $\alphacore$, $\bracket{\delta v^2}_{\rm mid}/c_s^2$, 
$\bracket{\delta \rho^2}_{\rm mid}/\bracket{\rho}_{\rm mid}^2$, and 
$\dot{m}_w/\bracket{\rho}_{\rm mid}c_s$. 
The horizontal axis shows the number of 
grid cells per length $\sqrt{2}h$ in the vertical direction, $n_z$.
The value  $n_z = 20$ corresponds to the resolution adopted in this study.
}
\label{fig:resolution}
\end{figure}
We carry out X1 simulations with changing the numerical resolution 
to $20\times 40\times 100$ cells and $80\times 160\times 400$ cells.
Figure~\ref{fig:resolution} compares the saturated values of various quantities 
obtained from the two runs with the values from the original X1 run (Table~2).
Here, the horizontal axis shows  
the number of grid cells per length $\sqrt{2}h$ in the vertical direction, $n_z$ 
(see footnote~\ref{foot}).
The value $n_z=20$ corresponds to our original resolution.
We see that the change of the resolution hardly affects 
the integrated accretion stresses $\alpha$ and $\alphacore$ 
and outflow flux $\dot{m}_w$, suggesting that the resolution of $n_z = 20$ 
is sufficient for these quantities to converge well.
By contrast, the velocity and density dispersions $\bracket{\delta v^2}_{\rm mid}$
and $\bracket{\delta \rho^2}_{\rm mid}$ increase with improving the resolution.
Since the energy input rate to turbulence should be the same  
if the integrated accretion stress is unchanged, 
the resolution dependence of the velocity and density dispersions
is expected to mainly come from artificial dissipation of sound waves in the simulation box.
However, we also see that this effect becomes less significant as the resolution is improved.
Detailed inspection shows that the fractional increase in $\bracket{\delta v^2}_{\rm mid}$ 
is $81\%$ when going from $n_z = 10$ to $n_z = 20$
but is $40\%$ when going from $n_z = 20$ to $n_z = 40$.
This suggests that the amplitudes of the velocity and density fluctuations 
should converge to finite values in the limit of high resolutions ($n_z\to\infty$). 
This is to be expected, since 
sound waves in a stratified disk physically  dissipate through, e.g., shock formation,
particularly at high altitudes where the gas density is low and therefore 
the amplitudes of the waves become large.
We find that the data for $\bracket{\delta v^2}_{\rm mid}/c_s^2$ 
shown in Figure~\ref{fig:resolution} lie on a curve 
$\bracket{\delta v^2}_{\rm mid}/c_s^2 = 0.012 - 0.0015 n_z^{-0.15}$.
This implies a converged value of $\bracket{\delta v^2}_{\rm mid}/c_s^2 \approx 0.012$, 
which is five times higher than that obtained in our $n_z = 20$ simulation
($\bracket{\delta v^2}_{\rm mid}/c_s^2 \approx 0.0024$). 
From this estimate, we see that $\bracket{\delta v^2}$ and $\bracket{\delta \rho^2}$ could be
underestimated by a factor of several in the simulations presented in this study.

In summary, we find that the outflow mass flux and vertically integrated accretion stress
converge well within our numerical resolution. This suggests that the predictor functions 
for $\alphacore$ and $\alphaatm$ (Equations~\eqref{eq:pred_core} and \eqref{eq:pred_atm}) 
and the scaling relation between $\dot{m}_w$ and $\alphaatm$ are hardly
affected by the resolution.
On the other hand, the amplitudes of sound waves could be underestimated 
by a factor of several because of the finite grid size.
Future high-resolution simulations will enable to better
quantify the scaling relations between $\bracket{\delta v^2}$  and $\alphacore$ 
and between $\bracket{\delta \rho^2}$ and $\alphacore$ 
(Equations~\eqref{eq:dv2} and \eqref{eq:drho2}).

\section{Summary}\label{sec:8}
Good knowledge about the turbulent structure of protoplanetary disks
is essential for understanding planet formation.
To provide an empirical basis for modeling the coevolution of dust and MRI,
we have performed MHD simulations of a vertically stratified shearing box 
with an MRI-inactive ``dead zone'' of various sizes and 
with a vertical magnetic flux of various strengths.
Our findings are summarized as follows.
\begin{enumerate}

\item
We have introduced the critical heights ($h_\ideal$, $h_\Lambda$, and $h_\res$)
that characterize the MRI in a stratified disk (Section~\ref{sec:3}). 
We have found that the vertical structure 
of MRI-driven turbulence depends on the resistivity profile 
only through the critical heights for the dead zone 
($h_\Lambda$ and $h_\res$) and is insensitive to the detail of the resistivity profile 
(Section~\ref{sec:4.2}).

\item 
In the ``disk core'' ($|z| < h_\ideal$), the density-weighted velocity dispersion 
$\rhodvtwo$ is nearly constant along the vertical direction (Section~\ref{sec:4.1}). 
This means that the velocity dispersion 
is approximately inversely proportional to the gas density.
Weak dependence on $z$ is also found for 
$\bracket{\delta\rho^2}/\bracket{\rho}$,
meaning that the density fluctuation $\bracket{\delta\rho^2}^{1/2}$ is proportional 
to the square root of the averaged density.

\item
The accretion stresses in the disk core and ``atmosphere'' ($|z| > h_\ideal$)
differently contribute to the turbulent structure of a disk (Section~\ref{sec:5.1}).
The velocity dispersion $\bracket{\delta v^2}$ 
and density fluctuation $\bracket{\delta \rho^2}$  in the disk core 
depend linearly on the accretion stress  integrated over the core, $\alphacore$ 
(Equations~\eqref{eq:dv2mid} and \eqref{eq:drho2mid}). 
By contrast, the outflow mass flux $\dot{m}_w$ 
depends linearly on the stress integrated over the atmosphere, 
$\alphaatm$ (Equation~\eqref{eq:wind}). 

\item
We have obtained simple empirical formulae 
that predict the vertically integrated stresses $\alphacore$ and $\alphaatm$ 
in the saturated state (Section~\ref{sec:5.2}; Equations~\eqref{eq:pred_core} and \eqref{eq:pred_atm}).
These are written as a function of the strength of the vertical magnetic flux (or $\beta_{z0}$)
and the critical heights of the dead zone 
measured in the nonturbulent state ($h_{\res,0}$ and $h_{\Lambda,0}$).
These predictor functions together with the saturation relations 
described above allow to calculate various turbulent quantities 
for a given resistivity profile and a net vertical flux.

\item
We have confirmed that the vertical diffusion coefficient $D_z$ of contaminants
is given by $D_z \approx \bracket{ \delta v_z^2}/\Omega$ 
both inside and outside a dead zone (Section~\ref{sec:6}).
This implies that sound waves propagating across a dead zone contribute to
the diffusion of dust particles just as turbulence does in active zones.
We have obtained a simple analytic recipe for the vertical distribution of $D_z$
as a function of $\alphacore$
on the basis of our MHD simulation data (Equation~\eqref{eq:Dz}).
\end{enumerate}

The empirical formulae obtained in this study enable us to predict  
the amplitudes of various turbulent quantities
in a protoplanetary disk with a dead zone.
The steps to be performed are as follows.

\begin{enumerate}
\item Prepare the vertical profile of the ohmic resistivity $\eta$, 
and find $h_\Lambda$ and $h_\res$ 
from Equations~\eqref{eq:h_Lambda} and \eqref{eq:h_res}.
A realistic profile of $\eta$ in the presence of dust particles
can be obtained by solving the ionization state of the gas and the charge state of the dust 
simultaneously \citep[e.g.,][]{SMUN00,IN06a,O09}.

\item Calculate $\alphacore$ and $\alphaatm$ using the predictor functions,
Equations \eqref{eq:pred_core} and \eqref{eq:pred_atm}.

\item One can now calculate the turbulent viscosity of the disk as 
$\nu_{\rm turb} = (3/2)(\alphacore+\alphaatm)c_s^2/\Omega$
(see Equations~\eqref{eq:alpha}, \eqref{eq:alphaint}, and \eqref{eq:alphaext}).
The vertical distribution of the gas velocity dispersion and density fluctuation 
in the disk core ($|z|<h_\ideal$) can be calculated 
from Equations~\eqref{eq:dv2} and \eqref{eq:drho2}, respectively.
The outflow mass flux can be evaluated from Equation~\eqref{eq:wind}.
For the diffusion coefficient in the disk core, one can use Equation~\eqref{eq:Dz}.
\end{enumerate}

When using our empirical formulae, 
it should be kept in mind that our scaling relations for velocity and density 
fluctuations (Equation~\eqref{eq:dv2} and \eqref{eq:drho2}) could
underestimate their mean-squared amplitudes relative to the integrated accretion stress
by a factor of several because of the numerical dissipation of sound waves (Section~\ref{sec:7}).
Future high-resolution simulations will allow to better quantify the saturation level 
of sound wave amplitudes. 

\acknowledgments
We are grateful to Neal Turner and Takayoshi Sano for providing 
us with their MHD simulation data that have motivated us to start this study.
We also thank Shu-ichiro Inutsuka, Takeru Suzuki, Taku Takeuchi, Hidekazu Tanaka, 
Takayuki Tanigawa, Mordecai-Mark Mac Low, and the anonymous referee
for useful discussion and fruitful comments. 
Calculations were made on the Cray XT4 at the CfCA, National Astronomical Observatory of Japan, 
and SR16000 at the Yukawa Institute for Theoretical Physics, Kyoto University.
S.O. is supported by a Grant-in-Aid for JSPS Fellows ($22\cdot 7006$) from the MEXT of Japan.

\appendix
\section{Ionization Degree and Ohmic Resistivity in Protoplanetary Disks}
In this Appendix, we explain how  
the resistivity profile adopted in this study (Equation~\eqref{eq:eta}) 
is related to realistic resistivity profiles in protoplanetary disks.
Since the resistivity is inversely proportional to the ionization degree 
(more precisely, the electron abundance; see \citealt*{BB94}), 
we will see how the ionization degree depends on the height $z$ above the midplane. 

Recombination occurs in the gas phase and on dust surfaces.
The gas-phase recombination dominates 
if the total surface area of dust particles is negligibly small.
In this case, the equation for the ionization-recombination equilibrium is given by  
\beq
\zeta n_n = \gamma_{ie} n_i n_e = \gamma_{ie} n_e^2,
\label{eq:equil_gas}
\eeq
where $\zeta$ is the ionization rate (the probability per unit time at which a molecule is ionized),
$n_n$, $n_i$ , and $n_e$ are the number densities of neutrals, ions, and electrons, respectively,
and $\gamma_{ie}$ is the gas-phase recombination rate coefficient.
The second equality in the above equation 
assumes the charge neutrality in the gas phase, $n_i = n_e$.
Equation~\eqref{eq:equil_gas} leads to the electron abundance
\beq
x_e = \frac{n_e}{n_n} = \sqrt{\frac{\zeta}{\gamma_{ie} n_n}} \propto \sqrt{\zeta}\exp\pfrac{z^2}{4h^2},
\eeq
which means that the resistivity is proportional to $\zeta^{-1/2}\exp(-z^2/4h^2)$.
Thus, if the vertical dependence of $\zeta$ can be neglected, 
the resistivity profile is given by Equation~\eqref{eq:eta} with $h_\eta =\sqrt{2}h$.
Note that Equation~\eqref{eq:eta_FS03} is derived instead of Equation~\eqref{eq:eta} 
if cosmic-ray ionization is assumed and the attenuation of cosmic rays toward the midplane
is taken into account \citep{FS03}.

If the total surface area of dust particles is large, recombination occurs mainly on dust surfaces.
In this case, the equation for the ionization-recombination equilibrium is given by
\beq
\zeta n_n = \gamma_{de} n_dn_e,
\label{eq:equil_dust}
\eeq
where  $\gamma_{de}$ is the sticking rate coefficient for dust--electron collision
and $n_d$ is the number density of dust particles.
The sticking rate coefficient depends on the charge of the dust particles,
and ultimately on $n_e$ via the charge neutrality \citep[see][]{O09},
but we will ignore this dependence in the following. 
From Equation~\eqref{eq:equil_dust} , we have
\beq
x_e = \frac{\zeta}{\gamma_{de}n_d} \propto \zeta\exp\pfrac{z^2}{2h_d^2},
\eeq
where we have assumed that $n_d \propto \exp(-z^2/h_d^2)$ 
with $h_d$ being the scale height of the dust particles
(in fact, one can show using Equation~\eqref{eq:advdiff} that 
$n_d$ obeys a Gaussian distribution in sedimentation-diffusion equilibrium 
if $D_z \propto \tau_s \propto n_g^{-1} \propto \exp(z^2/2h^2)$;
the condition $\tau_s \propto n_g^{-1}$ is satisfied 
if the size of the dust particles is smaller than the mean free path of the gas).
Thus, ignoring the dependence of $\zeta$ on $z$, 
the resistivity $\eta \propto x_e^{-1}$ is given by Equation~\eqref{eq:eta} with $h_\eta = h_d$.



\end{document}